\documentclass[12pt]{article}
\usepackage[margin=1in]{geometry}
\usepackage{amssymb,amsmath,amsthm,amsfonts,
float,eurosym,geometry,ulem,graphicx,caption,
setspace,sectsty,comment,footmisc,natbib,
pdflscape,subcaption,array, subfiles,adjustbox,threeparttable
}
\usepackage{dcolumn}
\usepackage{longtable, booktabs}
\usepackage{siunitx}
\usepackage{times}
\usepackage{hyperref}
\usepackage{enumitem}
\usepackage{tikz}
\usepackage{bbm}
\usepackage{rotating}
\usepackage{color}
\usepackage{pdfpages}

\usepackage{array}
\usepackage{multirow}
\usepackage{mathbbol}
\usepackage{accents}
\usepackage{authblk}
\usepackage{appendix}

\makeatletter
\g@addto@macro{\UrlBreaks}{\UrlOrds}
\makeatother

\newcommand{\OMIT}[1]{}

\newcommand\MyBox[2]{
  \fbox{\lower0.75cm
    \vbox to 1.7cm{\vfil
      \hbox to 1.7cm{\hfil\parbox{1.4cm}{#1\\#2}\hfil}
      \vfil}%
  }%
}
\definecolor{cblue}{rgb}{155, 221, 255}

\theoremstyle{definition}

\usepackage{xcolor}
\hypersetup{
    colorlinks,
    linkcolor={blue},
    citecolor={blue},
    urlcolor={black}
}

\newcolumntype{L}[1]{>{\raggedright\let\newline\\arraybackslash\hspace{0pt}}m{#1}}
\newcolumntype{C}[1]{>{\centering\let\newline\\arraybackslash\hspace{0pt}}m{#1}}
\newcolumntype{R}[1]{>{\raggedleft\let\newline\\arraybackslash\hspace{0pt}}m{#1}}

\newcommand{\email}[1]{\href{mailto:#1}{\nolinkurl{#1}}}

\begin{document}

\title{The Value of Personalized Recommendations: Evidence from Netflix\thanks{Corresponding authors: \email{kzielnicki@netflix.com},  \email{guy.aridor@kellogg.northwestern.edu}, and \email{abibaut@netflix.com}. Kevin, Aur\'elien, Allen, Winston, and Nathan were all employed and hold a financial stake in Netflix. Guy was a paid contractor at Netflix, but currently holds no financial stake in Netflix. We thank the editor \"{O}zlem Bedre Defolie and two anonymous reviewers whose comments greatly improved the paper. We thank Tommaso Bondi, Brett Gordon, Rafael Jim\'enez-Dur\'an, Malika Korganbekova, Olivia Natan, and Andrey Simonov as well as seminar/conference audiences at the ACM Conference on Economics and Computation, AI Algorithms and Markets Symposium (LUISS), Conference on AI and Human Decisions (CMU), Cornell Johnson, EARIE, NABE East Coast Tech Economists Meetup, NBER Digital Economics and AI Meeting, Netflix, Netflix Workshop on Personalization Recommendation and Search, and Research Roundtable on Platform Dynamics (Northwestern) for helpful comments. We thank Meghana Bhatt, Patric Glynn, Apoorva Lal, Adam Lindsay, Danielle Rich, Max Schmeiser, Qi Wang, and Patrick Zheng for internal assistance at Netflix with various aspects of this project. All errors are our own.}}

\author[1]{Kevin Zielnicki}
\author[2]{Guy Aridor}
\author[1]{Aur\'elien Bibaut}
\author[1]{Allen Tran}
\author[1]{Winston Chou}
\author[1,3]{Nathan Kallus}

\affil[1]{\small Netflix}
\affil[2]{\small Northwestern University, Kellogg School of Management}
\affil[3]{\small Cornell University and Cornell Tech}

\date{\today}
\maketitle
\thispagestyle{empty}
\begin{abstract}
Personalized recommendation systems shape much of user choice online, yet their targeted nature makes separating out the value of recommendation and the underlying goods challenging.  We build a discrete-choice model that embeds recommendation-induced utility, low-rank heterogeneity, and flexible state dependence and apply the model to viewership data at Netflix. We identify recommendation-induced engagement from observational variation in algorithmic exposure, treating exposure as conditionally exogenous given observed user, good, and state characteristics. Separately, we estimate model-free diversion ratios from a randomized experiment that perturbs recommendations and use these to assess the validity of the structural model. We use the model to evaluate counterfactuals that quantify the incremental engagement generated by personalized recommendations. First, we show that replacing the current recommender system with a matrix factorization or popularity-based algorithm would lead to a 4\% and 12\% reduction in engagement, respectively, and decreased consumption diversity. Second, most of the consumption increase from recommendations comes from effective targeting, not mechanical exposure, with the largest gains for mid-popularity goods (as opposed to broadly appealing or very niche goods).
\end{abstract}

\newpage

\onehalfspacing

\setcounter{page}{1}

\section{Introduction}

Online platforms increasingly mediate user choice through personalized recommendations, which alleviate the informational and search frictions of selecting high quality goods from large catalogs. These systems are deployed across a wide range of platforms in the digital economy, such as e-commerce and media streaming platforms. Their large role raises a fundamental question: how much of users' demand reflects the inherent demand for goods versus the effects of the recommendations? This distinction is important for quantifying the impact that the recommendation system (RecSys) has on platform engagement as well as quantifying the underlying incremental engagement generated by goods, which can be crucial for catalog optimization. However, this problem is challenging as recommendations are personalized to users' tastes, making it unclear whether a user consumed the good due to the recommendation or their underlying preferences.

In this paper, we estimate a discrete-choice model of user choice that incorporates the effects of recommendations on choice on a large video streaming platform with a prominent RecSys---Netflix. With this model in hand, we are able to both separately quantify the value of recommendations for engagement and provide a credible method for estimating the incremental demand for the underlying goods. A key aspect of our approach is to rely on natural and induced randomization in the recommendations in order to identify our model and to estimate model-free diversion ratios in order to validate our structural model.\footnote{Diversion ratios are a statistic used to quantify substitution patterns in industrial organization \citep{conlon2021empirical} that measure how a change in the quality or availability of one good changes the choice share of another.}

Modeling user choices in this setting faces several challenges that are present on many other online platforms. First, conditional on subscribing to the platform, users costlessly consume goods, meaning that there is no price variation that we can exploit to help us measure user valuation of goods or ascertain the role of recommendations. Second, estimating movie demand has historically been challenging as traditional good characteristics, such as cast and genre, have done a relatively poor job of predicting user demand \citep{einav2007seasonality, mckenzie2023economics}. Finally, there is potential serial dependence in choices (i.e., the good a user consumes today influences their choice tomorrow), and both the number of goods and users is incredibly large. We overcome each of these challenges and provide a model that we apply to estimate the value of goods and recommendations.

The structure of our model follows a low-rank discrete choice model \citep{kallus2016revealed}, which assumes that the good $\times$ user matrix of user preferences can be represented by a low-rank representation of good and user embeddings. We extend this baseline model in several ways to solve the aforementioned challenges. For the first challenge, we model the effects of recommendations via an additive recommendation ``bonus" that provides a utility boost for goods that are recommended. This is a reduced-form characterization of the effects of the recommendation as increasing the likelihood of consideration and providing idiosyncratic quality information for these goods \citep{aridor2022economics}. For the second challenge, we endogenously learn the good-level ``characteristics" directly from user choices, circumventing the need to define a reasonable set of characteristics that can rationalize choices. For the third challenge, we adapt a popular approach in the machine learning literature to our setting and consider a  transformer-style architecture \citep{vaswani2017attention} that extends standard state-dependent demand methods to allow for user preferences to flexibly adapt across time and rely on a user's consumption history.\footnote{The central concept of the transformer architecture is ``attention'', which is a context-dependent weight computed for each vector in a sequence, enabling a weighted-average summarization of an arbitrary-length sequence into a single vector. In this work, we use a simpler form of attention weighting described in Section \ref{sec:ccm}.}

We estimate our model using two million Netflix users and the approximately 7,000 goods available to them in the United States. At this scale, modeling users via pre-defined characteristics would be insufficiently expressive, while learning a latent preference vector for every user would be computationally infeasible. Instead, we represent each user’s preferences as a function of their past viewing history. Concretely, we use a model that reads a user’s sequence of previously watched goods (and their latent characteristics) and, at each point in time, summarizes this history into a current preference state. Thus, user heterogeneity is captured by a common set of parameters governing how histories map into preferences and the individual’s realized history, rather than by a user‑specific parameter vector. As a result, the number of parameters to estimate grows only with the number of goods and the complexity of the sequence model, rather than with the number of users. This dramatically reduces the memory and computational requirements for model estimation, enabling us to scale to large populations while capturing heterogeneous and time-varying user preferences.

To separate out the effects of recommendations from the intrinsic preferences of users, we treat algorithmic variation in the selection and ordering of recommended goods as conditionally exogenous given the user, item, and state features available to the RecSys. For validation, we conduct a randomized salience experiment that increases or decreases the share of recommendations from different product categories across consumers. This variation allows us to estimate model-free diversion ratios \citep{conlon2021empirical}. We generate the same diversion ratio estimates using our model and find that, out of sample, the model's predicted diversion ratios have a 0.86 correlation and 0.73 $R^{2}$ when compared to the model-free diversion ratios. This approach builds upon a recent line of work in industrial organization that relies on exogenous non-price variation, such as good availability experiments, in order to estimate diversion ratios in settings without price variation \citep{conlon2013experimental, conlon2023estimating, aridor2025measuring}. 

We use our model to better understand to what extent recommendations influence user choice. There are two distinct perspectives on how the RecSys affects choices. First, the RecSys can reduce the informational and search costs associated with choosing satisfying items from the catalog, thus increasing \textit{overall} consumption. Second, the personalization capabilities of the RecSys can especially help to enable consumption of niche goods, and thus have heterogeneous impacts across different types of goods. Motivated by this, we use our model to quantify how much engagement and consumption diversity is driven by the current RecSys compared to reasonable algorithmic benchmarks and quantify the role of effective targeting in driving consumption across the distribution of goods. We focus on overall engagement and diversity as these are known to correlate strongly with long-term customer satisfaction and retention, and are thus strong proxies for the value of personalization to the firm \citep{anderson2020algorithmic, bibaut2024learning, wang2022surrogate}. Although increased retention should also reflect greater consumer welfare, different underlying mechanisms would have different consumer welfare implications, and since our model utilizes a reduced-form formulation we refrain from making precise claims about the user value of personalized recommendations.

In order to quantify the overall value of the RecSys to the firm, we evaluate counterfactuals where the set of recommendations is generated by (i) randomized, (ii) popularity-based, and (iii) traditional matrix factorization  \citep{koren2009matrix} recommendation systems. These different approaches represent meaningful benchmarks: random represents the baseline effect of any recommendations, popularity represents the best non-personalized algorithm, and, as discussed in \cite{gomez2015netflix}, a matrix factorization based RecSys was the algorithm used at Netflix 10 years prior to this work.  Our main finding is that the current RecSys balances high engagement with high consumption diversity. Moving to an entirely random recommendation policy decreases engagement by 16\%, with minimal improvements to overall consumption diversity. Moving to a popularity-based policy or a traditional matrix factorization approach decreases engagement by 12\% and 4\%, respectively, while having large negative effects on consumption diversity. Internal estimates of the value of engagement suggest that these are economically large changes \citep[cf.][]{gomez2015netflix}. Furthermore, the current RecSys dramatically improves consumption diversity relative to the latter benchmarks, which is important for long-run user satisfaction \citep{anderson2020algorithmic}.

How does the value of the recommendation vary across goods and how important is targeting in driving this? We consider the difference in the model-implied probability of consuming a good for a targeted user who is recommended a good and an average user who is counterfactually not. The difference can be driven by three components: baseline propensity to consume (selection), the base effect of recommending a good to any user (exposure), and the larger incremental change in consumption probability on users who receive the recommendations (targeting). We find that, for the observation-weighted average across goods, the base effect of recommending a good to any user (exposure) is relatively large---exposing a good to the average user triples its consumption probability. However, in our decomposition, exposure only accounts for 6.8\% of the difference, with selection accounting for 51.3\% and targeting for 41.9\%. The magnitude of this effect is dramatically large, with targeting being 7 times larger than the mechanical exposure effect in explaining the consumption differences, whereas under the counterfactual matrix factorization algorithm both of these channels are nearly equally as important. We explore heterogeneity across goods and find that targeting most benefits mid-sized goods that appeal to a sizeable but specific subset of users relative to broadly appealing or very niche goods.

Finally, apart from the value of recommendations, we discuss how the model can be used to measure the incremental engagement generated by particular goods. We define the \textit{incremental} engagement for a particular good as the difference between the engagement with the good available on the platform compared to it being removed. We can credibly simulate this counterfactual since we can compute choice probabilities and recommendations under different counterfactual catalogs. One key limitation of our baseline model is that, since the characteristics of goods are learned endogenously based on choices, we cannot compute these counterfactuals for new goods that are not available on the platform. Thus, we provide an extension of the model that relies on high-dimensional, pre-consumption good embeddings that are derived from baseline good characteristics and human labels about each of the goods. Put together, these different components enable the estimation of the incremental demand associated with both new and existing goods that is not biased by the influence of recommendations on consumption choices.

\noindent \textbf{Related Work.} Our paper provides both substantive and methodological contributions. Substantively, we contribute to the literature on the impact of online recommendations on user choices. Methodologically, we provide innovations in using good recommendations to estimate diversion ratios and include a more flexible characterization of state-dependent demand estimation models.

\noindent \textbf{The Impact of Online Recommendations.} We contribute to the literature that quantifies the effects of recommendations and rankings on choices. \citet{SenecalNantel2004}, \citet{DasDatarGargRajaram2007}, \citet{FreyneJacoviGuyGeyer2009}, \citet{ZhouKhemmaratGao2010}, \citet{claussen2019editor}, \citet{HoltzCarteretteChandarNazariCramerAral2020WP}, \citet{aridor2022economics}, and \citet{donnelly2024welfare} show that, compared to non-personalized benchmarks, recommendation systems increased consumption in hypothetical
choices in a lab experiment, Google News, a social network, a news website, YouTube, Spotify, MovieLens, and
Wayfair, respectively. These papers collectively provide evidence that personalized recommendations, relative to a non-personalized benchmark, meaningfully increase consumption.
There has subsequently been a significant amount of work measuring how personalization impacts both the user-level and aggregate-level diversity of the consumption distribution  \citep{FlederHosanagar2009MS, NguyenHuiHarperTerveenKonstan2014, BrynjolfssonHuSimester2011MS,HosanagarFlederLeeBuja2013MS, LeeHosanagar2019ISR, aridor2020deconstructing, HoltzCarteretteChandarNazariCramerAral2020WP, KorganbekovaZuber2023WP, CalvanoCalzolariDenicoloPastorello2022WP}. 

Our work contributes to the literature in several ways. First, we quantify the incremental value of recommendation on Netflix, a platform with a highly influential and economically significant RecSys \citep{bennett2007netflix}. By comparing not only to non-personalized benchmarks, but also a previous state of the art algorithm (e.g., matrix factorization \citep{koren2009matrix}), we quantify the value of advances in personalization technology in the past decade (e.g., relative to those discussed in \cite{gomez2015netflix}). We showcase how this leads to differences in overall engagement as well as its influence on consumption diversity, which is especially important in large catalogs. Second, by systematically quantifying the value of recommendation across the good distribution and isolating how much of this comes from effective targeting, we more comprehensively document the heterogeneity in the value of recommendations across different goods---highlighting that mid-sized goods benefit the most and not goods with broad appeal or very niche goods (in contrast to the hypothesis of \cite{anderson2006long}).\footnote{This provides a complementary approach to \cite{lee2021recommender} who provide a theoretical model of recommendations that decomposes the value of recommendations into customization, screening, and selection. Our decomposition differs by focusing on the empirical forces that drive consumer choices.} Third, similar to the insights of \cite{morozov2021estimation} in other markets with search and information frictions, we highlight the importance of accounting for the influence of recommendations when estimating user preferences.

\noindent \textbf{Measuring Substitution Patterns.} Methodologically, our paper contributes to the literature on measuring user substitution patterns.

First, we contribute to the growing literature on measuring substitution patterns in markets with limited price variation. While these methods are broadly applicable, they are particularly relevant in the ``attention economy," where the scarce resource is user attention and goods are typically costless to consume \citep{brynjolfsson2019using, calvano2021market, yuan2025competing}.  Existing work relies on generated or exogenous good unavailability variation \citep{conlon2013experimental, raval2022using, aridor2025measuring} or using survey-based methods to elicit choices under hypothetical scenarios \citep{dertwinkel2024defining, bursztyn2025measuring}
to estimate second-choice diversion ratios. Facing a similar challenge, we highlight that platforms can use exploration experiments in the RecSys as plausibly exogenous variation to estimate both choice models and model-free diversion ratios.

Second, we contribute to the literature on state-dependent demand estimation, which provides a class of models to rationalize persistence in user choices. The empirical challenge in this literature, dating back to \cite{heckman1981heterogeneity, heckman1991identifying}, is to disentangle whether past choices causally affect future choices or if choices are spuriously serially correlated due to unobserved preference heterogeneity. The canonical models of state-dependent demand \citep{dube2010state, simonov2020state} provide a Markovian formulation with rich preference heterogeneity to solve this problem. We formulate our problem in a similar manner, except in our context we are more interested in how to model the evolution of preferences over time as opposed to decoupling these two channels. In order to do so, we adapt methods from transformer-style architectures that are popular in machine learning \citep{vaswani2017attention} to characterize state-dependence based on the full consumption history. This allows users' tastes to flexibly change over time and computationally simplifies our model by requiring us to only keep track of different consumption histories, instead of the full matrix of user embeddings. The latter enables us to estimate this model at the scale of millions of users with minimal compute requirements. Topically, \cite{zeller20222drip, lee2025pay} apply models with state dependence in media streaming contexts, albeit with substantively different models than we consider here.

\section{Institutional Details}\label{sec:institutional_details}

\begin{figure}[ht]
\centering
\caption{Netflix Homepage}
\label{fig:netflix_homepage}
\includegraphics[width=0.8\textwidth]{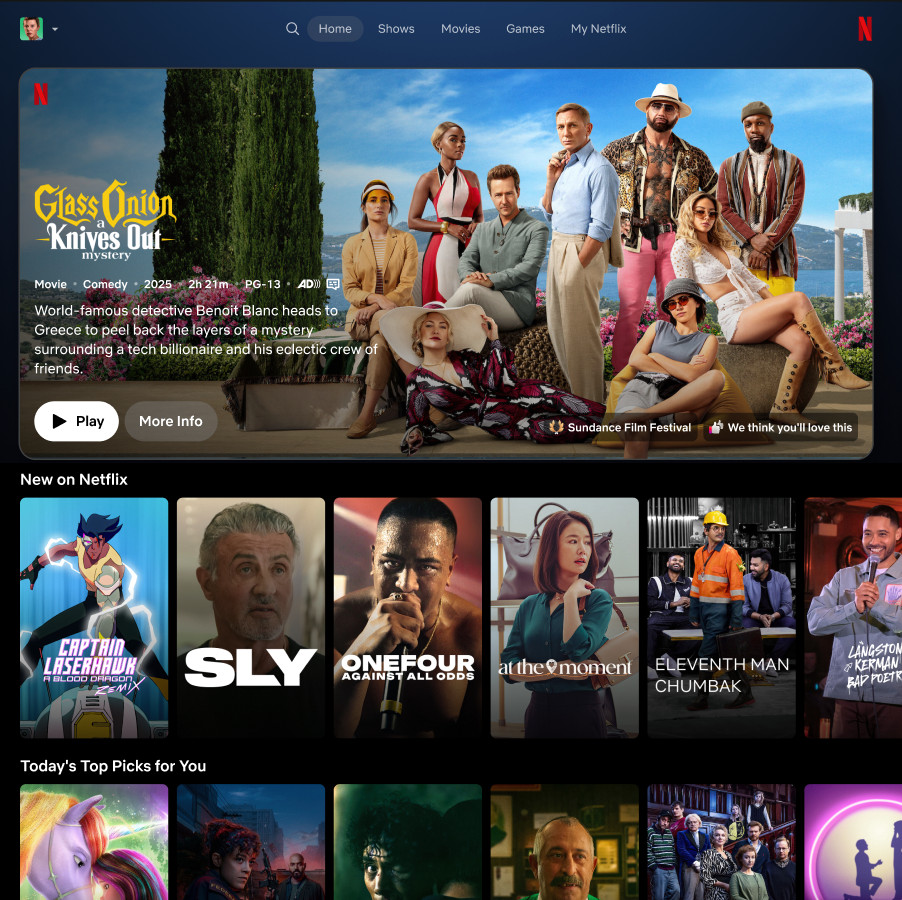}
\end{figure}

Our empirical context is the video-streaming platform Netflix. Users pay a subscription fee to access Netflix and can then costlessly consume any good they wish on the platform.  The set of goods consists of both television shows (which have multiple episodes and seasons) and movies. The Netflix catalog is vast, spanning thousands of goods, and we focus our empirical analysis on the United States, which has approximately 7,000 goods available as of October 2025.

To help users find satisfying goods, Netflix's homepage is personalized. There has been considerable development of the RecSys, dating back to the public Netflix prize \citep{bennett2007netflix} and disclosed in \cite{gomez2015netflix}. The implementation discussed in \cite{gomez2015netflix} relies on matrix factorization approaches, but since then the RecSys has evolved to incorporate newer algorithmic advances in generating recommendations (e.g., see \cite{steck2021deep}, \cite{Lamkhede2024}, \cite{hsiao2025foundation}).

Figure \ref{fig:netflix_homepage} provides an example of the homepage for one particular user. There are several distinct components of the homepage, which are important for us to consider when modeling it. First, there is one good that is highlighted for each user that spans across the width of the screen, which we will refer to as the billboard. Second, the example in Figure \ref{fig:netflix_homepage} shows that the recommendations are arranged in different rows that have a similar style grouping of goods, such as New on Netflix and Top Picks for You in this example. Based on this, the top-left $5 \times 5$ block of titles (i.e., the first 5 rows and the first 5 goods in each row) is where most users engage with the recommendations and we refer to these as the ``top 25" recommendations. Third, expanding to include the top-left $10 \times 10$ block of titles (i.e., the first 10 rows and the first 10 goods in each row), is where the vast majority of engagement from recommendations occurs and we refer to these as the ``top 100" recommendations. In our empirical model, we consider each of these canvases separately. Another important aspect of the homepage is that it includes a ``continue watching" row for users to continue consumption of a previously consumed good (e.g., an unfinished movie or a continuing TV show), which we do not consider a recommendation and is unimpacted by our interventions and counterfactuals.

\section{Demand Model}\label{sec:ccm}

We consider a user $i$ who chooses to consume a good from $\mathcal{M} \cup \{ 0 \}$ where $\mathcal{M}$ denotes the set of $J$ goods that are present on the platform and $0$ denotes the outside option (not consuming any good). In principle, within a given time period (for instance, a day), users can choose several goods and allocate continuous amounts of time to each of them. However, to capture the consumption of the modal user, we cast this problem as a discrete choice problem where a user chooses to consume a single good from $\mathcal{M}$ or take up the outside option at the time granularity of one day.\footnote{We consider consumption of a good as whether a user watches the good for a prespecified number of minutes. In the cases when a user watches multiple times in a given day, we randomly choose among those goods. We make this assumption since it simplifies the problem dramatically and empirically captures most viewership.} We consider all consumption on the platform, including both consumption originating from the recommendations and manual search.

\noindent \textbf{Model Formulation.} Given this formulation, we consider the following utility specification for user $i$ and good $j$ at time $t$:
\begin{align}\label{eq:utility}
u_{ijt} =
A_{it} B_j^\top + \sum_{r \in \{\text{billboard, top 25, top 100} \}}\beta_{jr} \mathbbm{1}\{j \in \mathcal{C}_{irt}\} + \varepsilon_{ijt} 
\end{align}
and we normalize the utility of the outside option $u_{i0t} = \varepsilon_{i0t}$.  Again, the outside option here denotes not consuming any good on Netflix on a given day.

This utility formulation is intrinsically in good space as it considers that our goal is to learn the matrix of user $\times$ good utilities. We follow a similar formulation as \cite{kallus2016revealed} by assuming that this can be decomposed into low-rank good and user matrices. We exogenously impose the rank of these matrices, such that $B$ is $\mathbb{R}^{J \times d}$ and $A_{t}$ is $\mathbb{R}^{I \times d}$. $A_{t}$ represents the user preference weights for each of the learned attributes at time $t$ and $B$ represents the endogenously learned characteristic space.\footnote{Note that both the ``preference" vector ($A_{t}$) and ``good characteristics" ($B$) are learnable parameters of the model, as opposed to one being a coefficient vector multiplying a vector of predictors, as is commonly the case in demand modeling, in which case would the second would be a random variable. In particular, there is no notion of ``correlation'' across time points of the learned title characteristics, and therefore no simultaneity / endogeneity concern. We refer the interested reader to \cite{kallus2016revealed} for a thorough analysis of joint learning of characteristics and preferences under low rank constraints.} Beyond these two factors, which represent the ``intrinsic" preferences that users have for the goods, we explicitly model the role of recommendations where $\mathcal{C}_{it} \equiv \cup_{r \in \{\text{billboard, top 25, top 100} \}} \mathcal{C}_{irt}$ denotes the set of recommendations that user $i$ observes at time $t$. We handle the granularity of the different types of recommendations (as described in Section \ref{sec:institutional_details}) by having separate recommendation utility booster terms for the billboard, top 25, and top 100 recommendations (denoted by $\beta_{jr}$). Finally, $\varepsilon_{ijt}$ denotes the standard Type-1 Extreme Value error. Under the standard assumption that $\varepsilon_{ijt}$ are independent and identically distributed, this induces the probability that good $j$ will be chosen by user $i$ in time period $t$:
\begin{align*}
\Pr(\text{i chooses j at time t}) =
\frac{
\exp\!\left(A_{it} B_j^\top + \sum_{r \in \{\text{billboard, top 25, top 100}\}} \beta_{jr} \mathbbm{1}\{j \in \mathcal{C}_{irt}\}\right)
}{
1 + \sum_{k \in \mathcal{M}}
\exp\!\left(A_{it} B_k^\top + \sum_{r \in \{\text{billboard, top 25, top 100}\}} \beta_{kr} \mathbbm{1}\{k \in \mathcal{C}_{irt}\}\right)
}
\end{align*}

A particularly important quantity for our counterfactuals and applications of the model is whether a user chooses any good. Thus, we define whether a user $i$ \textit{engages} in a time period $t$ as follows:
\begin{align}\label{eq:engagement}
    \textsc{engagement}_{it}(\mathcal{M}) = \sum\limits_{m \in \mathcal{M}} \Pr(\textrm{i chooses m at time t}) = 1 - \Pr(\textrm{i chooses $0$ at time t})
\end{align}
\noindent \textbf{Model Challenges and Assumptions.} There are three key challenges for credibly modeling user choices that we tackle with the structure of the choice model.

The first challenge is that, in our empirical context, recommendations play a large role in driving user choices and movies and TV shows are experience goods.\footnote{Because the goods are experience goods, users only learn their true utility after consumption, even with extensive search. By observing other users’ consumption, the RecSys can reveal idiosyncratic match value in addition to information about observable product characteristics.} Following \cite{aridor2022economics}, recommendations can influence user choices through two mechanisms: making users aware of goods they did not know existed (awareness/consideration) as well as providing information on product attributes and a signal of idiosyncratic quality (information). \cite{aridor2022economics} experimentally test for the relative importance of these two mechanisms on another movie recommendation platform -- MovieLens -- and find that the latter plays a meaningful role. As such, we consider a modeling approach that in principle approximates both mechanisms by modeling the effect of recommendation as providing an additive (ex ante) utility bonus for the set of goods that are recommended that captures its role in both shaping consideration and providing information.\footnote{Capturing both forces matters for rationalizing choices as a specification that only influences consideration sets (following \cite{yu2024welfare}) does not perform as well as our primary model according to our validation exercises.}$^{,}$\footnote{The recommendation bonus $\beta_{jr}$ is homogeneous across users, so the response
to a recommendation varies across users only through their intrinsic
preferences and choice sets. Our specification therefore approximates a consideration-set model as it does not capture user-specific responsiveness to recommendations and the counterfactuals in Section \ref{sec:rec_value} should be interpreted with this approximation in mind.}

The second challenge is that modeling demand for movies and TV shows has traditionally been difficult since traditional good characteristics (e.g., genre, cast, etc.) are typically not sufficient to explain user choices \citep{basuroy2003critical, einav2007seasonality, mckenzie2023economics}. Instead of relying on traditional characteristics, we endogenously learn the good embeddings via user choices in a similar spirit as traditional matrix factorization approaches used in recommendation systems \citep{mnih2007probabilistic, koren2009matrix}. As in \cite{kallus2016revealed}, the key difference is that we embed this within a discrete-choice model that allows us to flexibly learn these good embeddings. While this is our preferred specification, we nonetheless show that the model performs reasonably well if we use good-level embeddings that are learned on pre-consumption good characteristics and tags in Appendix \ref{sec:app_exog_embeddings}.

\begin{figure}
    \centering
    \caption{Preference Weight ($A_{it}$) Sequence Model}
    \label{fig:architecture}
    \includegraphics[width=0.8\linewidth]{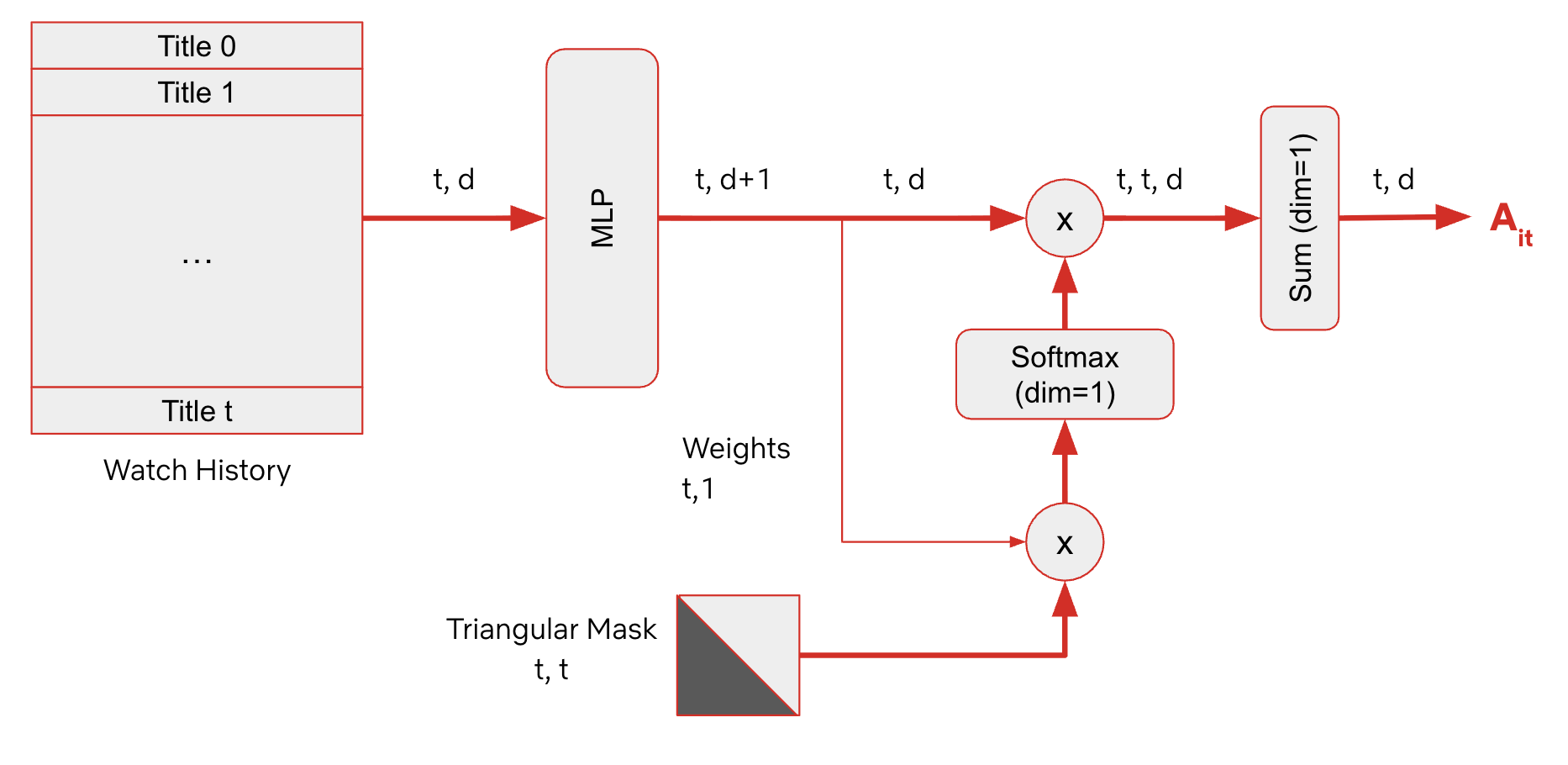}
    \caption*{\footnotesize\textsc{Notes}: This figure illustrates how time-dependent user preferences are derived from the user's sequence of chosen goods. First, we look up the good embeddings corresponding to the user's watch history. Then, we transform embeddings via a multi-layer perceptron, which also produces attention weights for each good. In order to efficiently parallelize across time points, we mask out any weights for goods chosen after that time point. Next, we normalize the non-masked attention weights at each time point with a softmax operation. Finally, we use the normalized weights to construct a weighted sum of the transformed good embeddings, representing the user's current preference state at each time point.}
\end{figure}

The third challenge is that there may be serial dependence in choices, for instance due to users being more likely to watch more sports-related shows after watching a sports-related movie. We let the preference weights, $A_{it}$, vary over time and encapsulate everything the user $i$ has watched up to period $t-1$, based on a sequence model of prior watch history as diagrammed in Figure \ref{fig:architecture}. At a given time $t$, we encode the $d$-dimensional good embeddings $B_{j}$ in an ordered sequence of when the user watched them. We pass that vector through a two-layer fully-connected multi-layer perceptron that produces, for each time period $t$, a set of attention weights for each previous time period and a transformed, user-specific, set of characteristics.\footnote{A two-layer fully connected multi-layer perceptron refers to a feedforward neural network with one hidden layer and one output layer, where each unit in one layer is connected to all units in the next \citep{goodfellow2016deep}.} Taking the weighted sum of these and passing it through a two-layer fully connected multi-layer perceptron again provides our vector of estimates $A_{it}$. Intuitively, the formulation is similar to a traditional sequence model in machine learning where the future goods that a user consumes depend on their previous sequence and the model endogenously learns position-specific attention weights. This aspect of the formulation is a modified variant of state-dependent demand models \citep{dube2010state} that combines both time-variation in preferences and ``intrinsic" preferences. In addition to addressing serial dependence, this architecture avoids the need to learn user embeddings directly, which becomes highly computationally challenging at the scale of millions of users.

\noindent \textbf{Identification.} The main identification challenge  is that recommendations are endogenously targeted towards users whom the RecSys predicts will have high predicted utility for that good. Thus, it is unclear without exogenous variation whether the user consumed the good because of their intrinsic preferences or the recommendation. We overcome this challenge by conditioning on the same observed user, good, and state characteristics available to the RecSys and treating the residual variation in exposure as conditionally exogenous given these features.  This variation can be attributed to the random exploration that the RecSys induces for learning, which allows us to separate the two components of interest.\footnote{We find similar estimates for the recommendation bonuses and model validation statistics when instead of relying on the RecSys exploration we directly estimate the model over data from a subset of the randomized salience experiments discussed in Section~\ref{sec:good_salience_exp}.} It also enables us to identify the effect of state dependence, since users are more likely to consume goods that appear in their recommendations.\footnote{Additionally, the same identification argument for the dynamic component of demand as in \cite{zeller20222drip} applies in our context. The identification argument is that, unlike in traditional models of state-dependent demand where the consumer can consume the same good repeatedly across periods, movies and TV shows have finite runtimes. Thus, at the end of the consumption the consumer ``exogenously" needs to find new goods to consume.}

\noindent \textbf{Computational Details for Estimation.} To estimate our model at scale, we employ stochastic gradient descent (SGD) to maximize the likelihood function, using mini-batch training implemented in PyTorch. Given the scale -- millions of users and thousands of goods --  mini-batch SGD is essential for both computational efficiency and tractability. We train our model on GPUs using AWS cloud infrastructure, which provides significant speedups over CPU-based computation. To manage the scale of the data, we preprocess user and viewing records using Spark. During training, batches of data are dynamically sampled and pre-fetched into memory, ensuring high GPU utilization and efficient learning.

\section{Model Estimates and Validation}

We estimate our model using the viewership data for a random sample of approximately 2 million active Netflix U.S. users for a period of 35 days (February 18, 2025 to March 25, 2025). We first provide the model estimates and then discuss our experimental validation.

\subsection{Model Estimates}

Since the model endogenously learns the characteristic space, we show in Figure \ref{fig:similarities} that the model produces reasonable good-level similarities. For instance, it considers ``Love Is Blind" most similar to other dating shows like ``Married at First Sight" but dissimilar to action films like ``Faster", and ``Despicable Me 4" most similar to other kids movies like ``Sonic the Hedgehog 2" but dissimilar to adult dramas like ``A Man in Full". 

\begin{figure}
\centering
    \caption{UMAP-projected Good Embeddings}\label{fig:goods}
    \begin{subfigure}[b]{0.5\textwidth}   
        \caption{Genre}         
        \includegraphics[width=\textwidth]{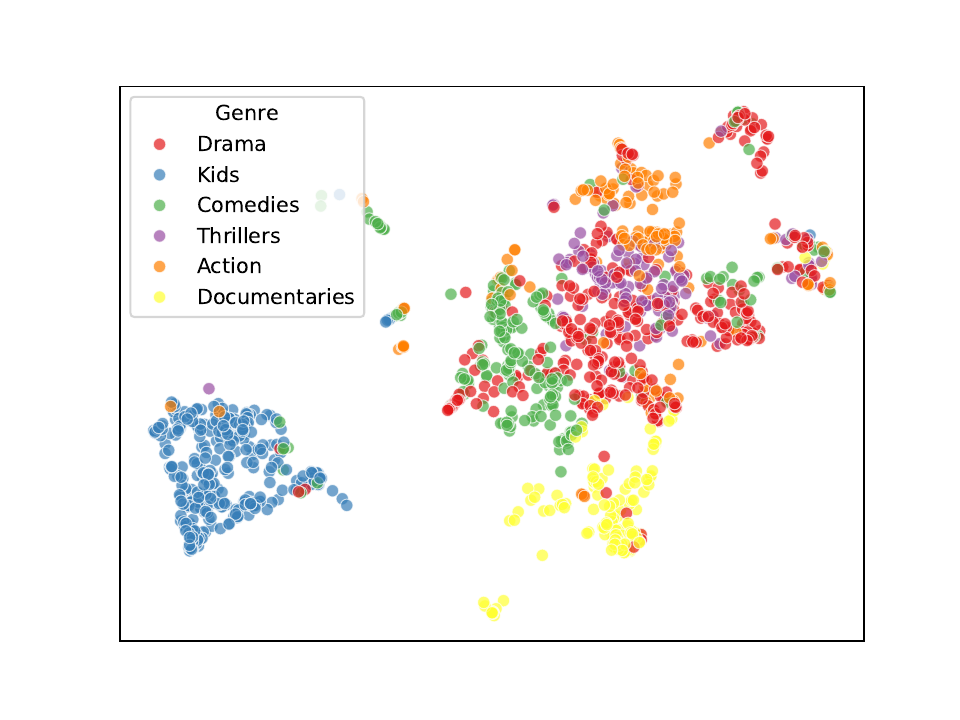}
        \label{fig:good_genre}
    \end{subfigure}%
    \begin{subfigure}[b]{0.5\textwidth}
        \centering
        \caption{Language}
        \includegraphics[width=\textwidth]{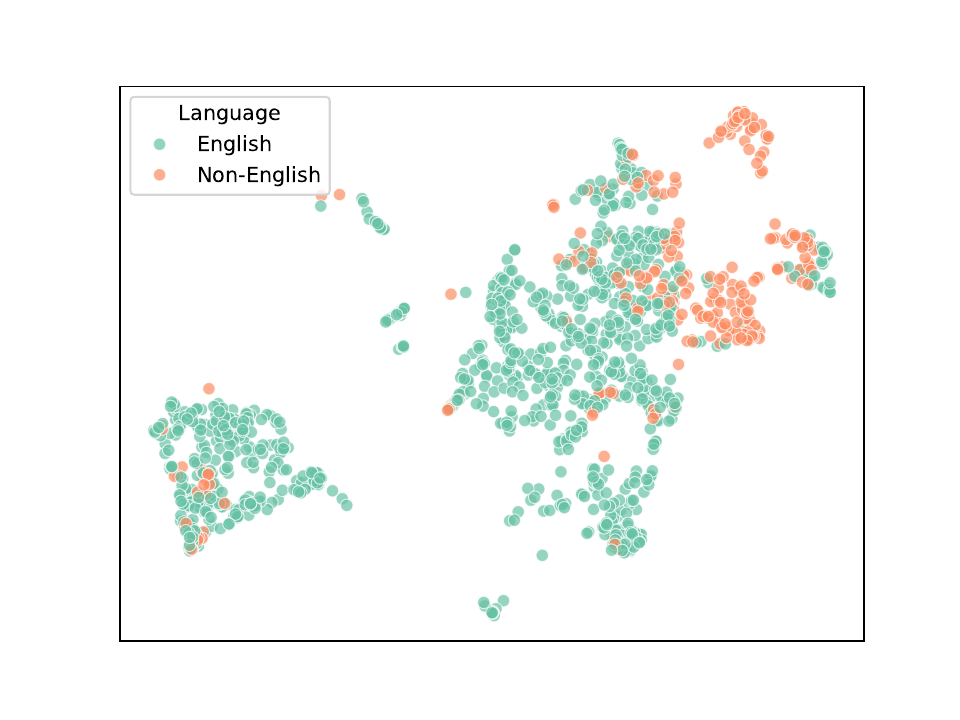}
        \label{fig:good_lang}
    \end{subfigure}
    \begin{subfigure}[b]{0.5\textwidth}   
        \caption{Movie vs. TV}         
        \includegraphics[width=\textwidth]{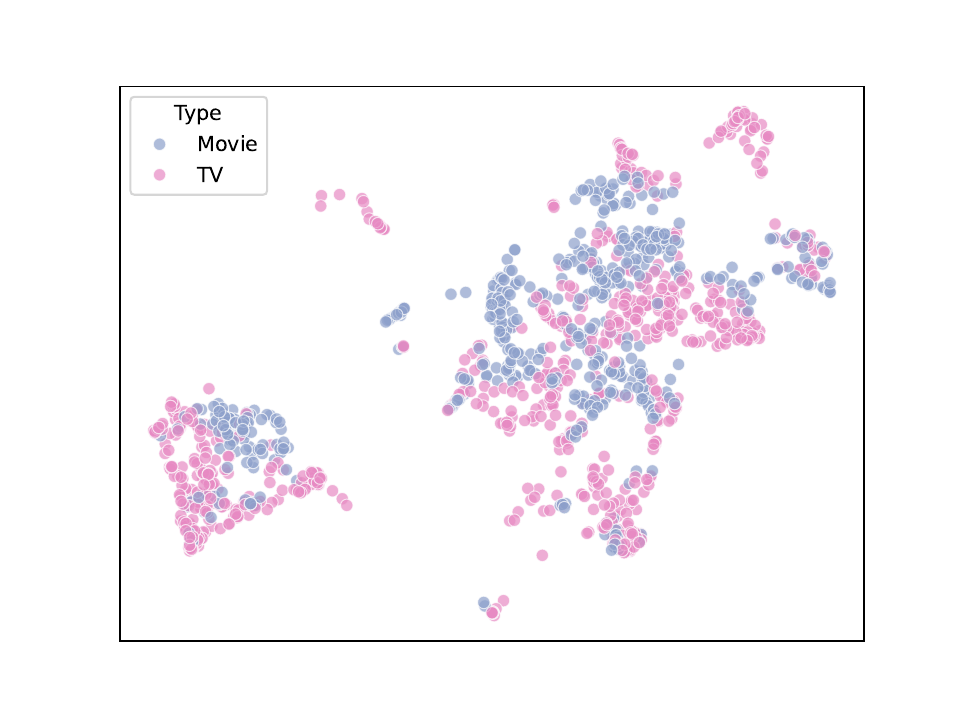}
        \label{fig:good_type}
    \end{subfigure}%
    \begin{subfigure}[b]{0.5\textwidth}
        \centering
        \caption{Popularity}
        \includegraphics[width=\textwidth]{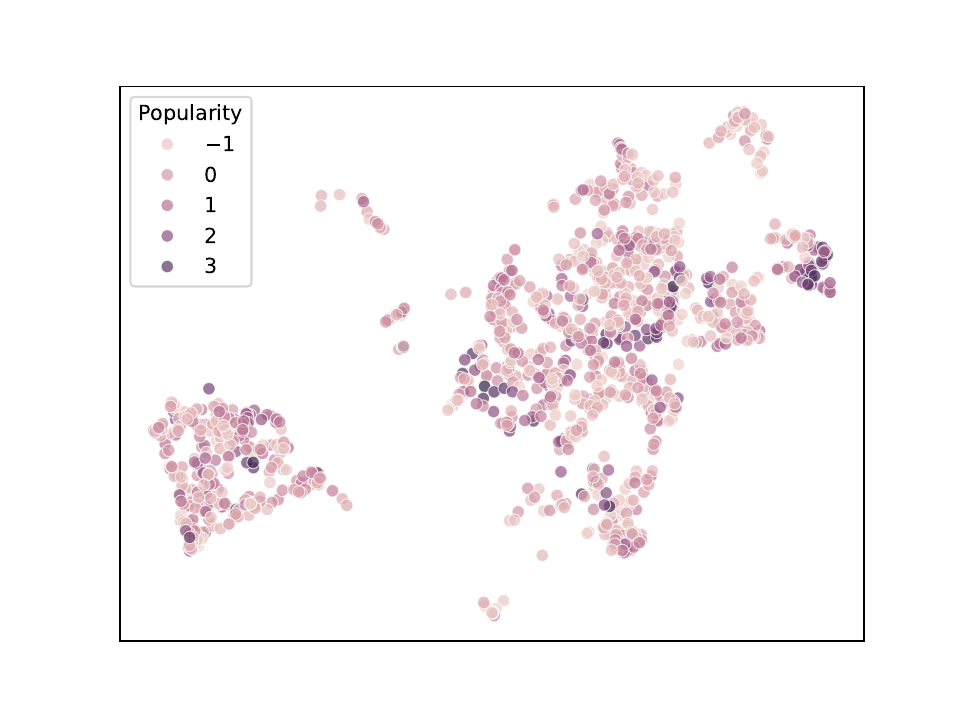}
        \label{fig:good_pop}
    \end{subfigure}
\caption*{\footnotesize\textsc{Notes}: These figures present two-dimensional UMAP projections of learned good embeddings $B_j$. Panels color points by (a) genre, (b) language, (c) movie vs.\ TV, and (d) popularity (share). Projections are for visualization only; all estimation is done in the original embedding space.}

\end{figure}

In Figure \ref{fig:goods} we show how a projection of the good embeddings via UMAP \citep{mcinnes2018umap-software} corresponds to characteristic good attributes. Figure \ref{fig:good_genre} shows that goods that are distinctly different from others (e.g., kids and documentaries content) are clustered together, while the remaining genres are similarly clustered together though there is considerable overlap with other genres.\footnote{This highlights the challenges with using observable characteristics, such as cast and genre, alone.} Additionally, Figure \ref{fig:good_lang} shows a distinct separation between non-English and English goods and Figure \ref{fig:good_type} highlights that, within clusters, there is a clear separation between movies and television shows. Overall, these figures highlight that the endogenously learned characteristic space is reasonable and accords with intuition, but also subtly captures more complex similarity between goods than can be discerned from standard observable characteristics.

\begin{figure}[ht]
\centering
\caption{Most and Least Similar Goods for Selected Goods}
\begin{tabular}{lll}
\toprule
\textbf{Good} & \textbf{Comparison Goods} & \textbf{Similarity} \\
\midrule
\multirow{5}{*}{Love Is Blind}
    & Love Is Blind: UK & 0.502 \\
    & Married at First Sight & 0.476 \\
    & The Millionaire Matchmaker & 0.470 \\
    & \textit{Red Notice} & -0.310 \\
    & \textit{Faster} & -0.356 \\
\midrule
\multirow{5}{*}{Zero Day}
    & To Catch a Killer & 0.690 \\
    & The Sum of All Fears & 0.559 \\
    & Trial by Fire & 0.499 \\
    & \textit{Total Drama} & -0.433 \\
    & \textit{That Girl Lay Lay} & -0.448 \\
\midrule
\multirow{5}{*}{American Murder: Gabby Petito}
    & The Search For Instagram's Worst Con Artist & 0.559 \\
    & Trial by Fire & 0.507 \\
    & American Murder: Laci Peterson & 0.487 \\
    & \textit{Robin Hood} & -0.359 \\
    & \textit{Wolf King} & -0.420 \\
\midrule
\multirow{5}{*}{Despicable Me 4}
    & Sonic the Hedgehog 2 & 0.726 \\
    & Plankton: The Movie & 0.709 \\
    & Minions & 0.598 \\
    & \textit{Gangs of London} & -0.448 \\
    & \textit{A Man in Full} & -0.525 \\
\bottomrule
\end{tabular}
\caption*{\footnotesize \textsc{Notes}: Similarity is cosine similarity between learned good embeddings $B_j$ from the baseline model estimated on U.S. viewing. Positive values indicate higher embedding similarity and negative values indicate relative dissimilarity. We show the three most and two least similar goods. The set of comparison goods is chosen for illustration and is not exhaustive. }

\label{fig:similarities}
\end{figure}

\subsection{Model Validation}\label{sec:model_validation}

We conduct several exercises to validate that our model accurately captures substitution patterns.

\begin{figure}
\centering      
    \caption{Observed Good Shares vs In-Sample Model  Estimates}\label{fig:shares}
    \includegraphics[width=0.5\textwidth]{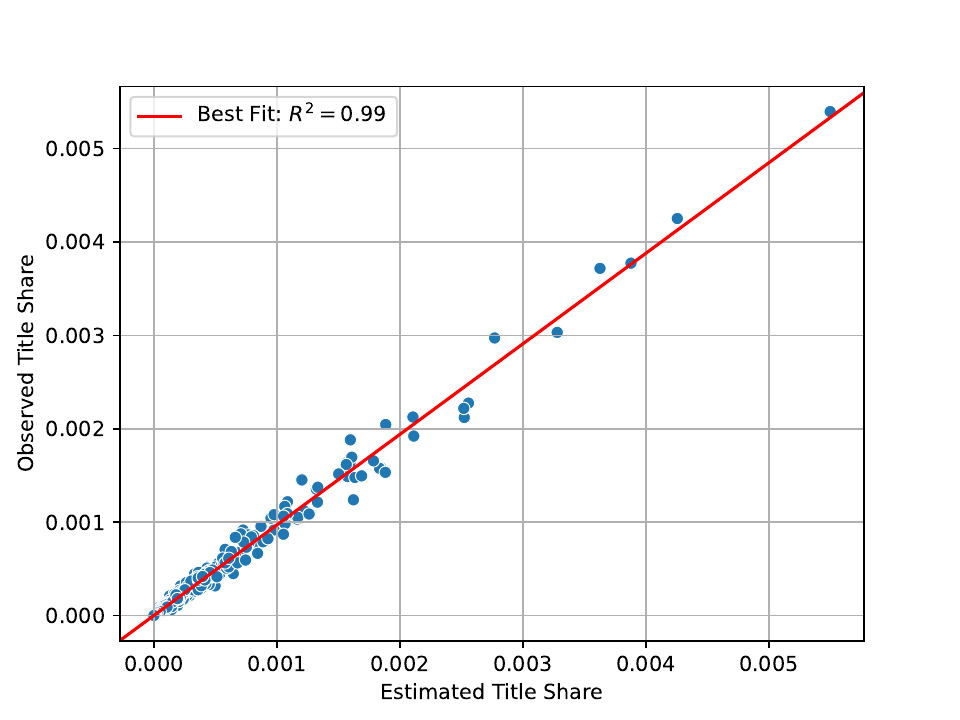}
    \caption*{\footnotesize\textsc{Notes}: This figure presents the comparison of the estimated and actual viewership shares for different goods. Each point is a good; $x$-axis shows observed share over the estimation window, $y$-axis shows model-implied in-sample share holding realized recommendations fixed.}

\end{figure}

One key challenge is that the consumption distribution follows a heavy-tail distribution, with some goods having an outsized share of consumption \citep{mckenzie2023economics}, and the challenge with traditional characteristics space models is that the good characteristics such as cast and genre cannot rationalize these patterns. Thus, our first simple validation exercise, presented in Figure \ref{fig:shares}, shows that  our model (in-sample) recovers good-level shares reasonably well.

Our next set of validation exercises rely on experimental modifications of the salience of particular goods in the recommendations in order to estimate model-free diversion ratios that we can benchmark against our model. We begin by describing the experiments that we conduct to generate these diversion ratios in Section \ref{sec:good_salience_exp} and then compare the model-based and experimental diversion rates in Section \ref{subsec:experimental_comparison}.

\subsubsection{Good Salience Experiments}\label{sec:good_salience_exp}

As discussed in Section \ref{sec:ccm}, we rely on idiosyncratic variation in recommendations for model identification. To validate that our model does in fact identify the causal effect of recommendations, we run a true randomized experiment that ``nudges'' recommendations exogenously.  In this experiment, we randomly allocate users into eight treatment arms corresponding to a focal good category (e.g., scripted vs. unscripted goods, language, film vs. television series).  Next, we substitute impressions of a random subset of goods not in the focal category for impressions of goods in the focal category.  Thus, in each treatment arm, we observe a set of focal goods whose salience is ``boosted'' (i.e., show up more in the recommendations) and a set of nonfocal goods whose salience is ``diluted'' (i.e., show up less in the recommendations).

The experiment is run for 5 weeks with each experimental arm including approximately 1 million users.  The resulting fluctuations in good salience are comparable with those in routine experiments that are run on the RecSys (e.g., to test algorithmic improvements).  Our experiment simply coordinates this variation across goods and users to generate more precise and interpretable model-free diversion ratios.

\subsubsection{Comparison to Experimental Estimates}\label{subsec:experimental_comparison}

\begin{figure}[H]
\caption{Modeled ($D_{g \to k}^{\text{MODEL}}$) vs. Empirical ($D_{g \to k}^{\text{EMPIRICAL}}$) Good Diversion}
\label{fig:model_validation}
\centering
    \includegraphics[width=0.8\textwidth]{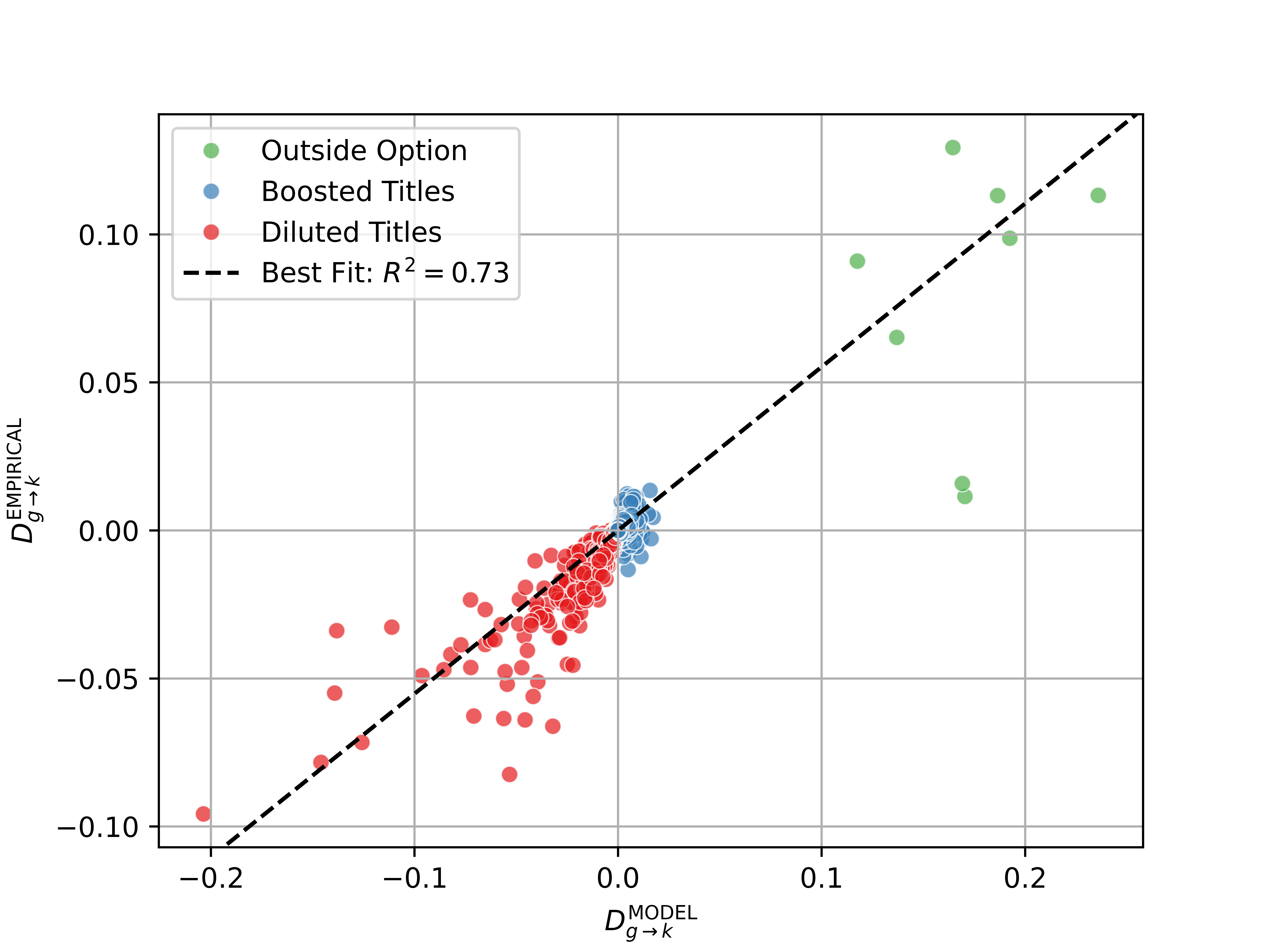}
\caption*{\footnotesize \textsc{Notes}: This figure presents comparison of diversion $D_{g\to k}$ computed from (i) experimental salience interventions (empirical) and (ii) model simulations that emulate the same recommendation restrictions (modeled). For the model-based diversion estimates, we estimate our model on the control arm and simulate the experimental intervention by imputing recommendations using the recommendation-model described in Appendix~\ref{sec:impute_recs}. The blue and red points denote goods whose recommendations were boosted and diluted, respectively. The green points denote diversion to the outside for each of the eight separate interventions.}

\end{figure}

We denote each of the treatment arms as $g \in \mathcal{G}$ and denote $P_{k}(g)$ as the probability that good $k$ gets chosen under the recommendation restriction $g$. We note that the goods in treatment arm $g$ represent the set of goods whose presence in the recommendations is diluted during the intervention. Therefore, following \cite{conlon2021empirical} we can use the Wald Estimator to obtain the main diversion measure of interest:
\begin{align}\label{eq:diversion_measure}
D_{g \to k} = \frac{P_k(g) - P_k(\emptyset)}{- \sum_{j \in g} \left( P_j(g) - P_j(\emptyset) \right)}
\end{align}
This provides the diversion to a specific good $k$ that results from experimental restriction $g$. We can therefore compare the performance of our model on these experimental measures of diversion in order to benchmark its performance in learning substitution patterns. We can easily estimate this quantity using the experimental data since $P_{k}(\emptyset)$ and $P_{j}(\emptyset)$ represent the empirical probability of choice for $k$ and $j$, respectively, in the control arm and $P_{k}(g)$ and $P_{j}(g)$ represent the empirical probability of choice, respectively, for $k$ and $j$ in treatment group $g$. We note that for $k = 0$ this captures the overall change in $\textsc{engagement}$ as a result of the intervention. We denote the experimental estimates as $D_{g \to k}^{\text{EMPIRICAL}}$.

In order to compare the performance of our model to these empirical quantities, we first estimate our model on the control arm. Then, we simulate each experimental group $g \in \mathcal{G}$ separately and produce the model estimates of $D_{g \to k}^{\text{MODEL}}$. Our main model validation test will then be to compare $D_{g \to k}^{\text{MODEL}}$ to $D_{g \to k}^{\text{EMPIRICAL}}$. In order to perform this validation exercise, we need to emulate the good salience intervention on the control arm. This requires us to define a procedure that replaces goods that appear in the recommendations with the goods that the RecSys would replace them with. In Appendix \ref{sec:impute_recs}, we provide two imputation procedures and validate them on the treatment arms of the good salience experiments. For the validation exercise, we use the recommendation model described in Appendix \ref{sec:impute_recs}, which leads to an $R^2=0.69$ when benchmarked against the actual recommendations generated during the salience experiment.

In Figure \ref{fig:model_validation}, we compare the estimated diversion ratios between the model ($D_{g \to k}^{\text{MODEL}}$)  and the experimental estimates ($D_{g \to k}^{\text{EMPIRICAL}}$). The figure shows the diversion to goods that were ``boosted" in blue, to goods that were ``diluted'' in the treatment groups $\mathcal{G}$ in red, and to the outside option in green. We find that the association between the counterfactual diversion ratios is strong with a correlation of 0.86 and an $R^{2}$ of 0.73. These results show that our model is able to reproduce out-of-sample estimates of both good-level diversion as well as diversion to the outside option, indicating its ability to capture substitution patterns.

\section{Estimating the Value of Recommendations and Goods}\label{sec:rec_value}

In this section, we describe two applications of our model: first, we quantify the value of recommendations in terms of the incremental engagement they generate; second, we illustrate how to use the model to quantify the incremental demand for goods.

\subsection{Value of Recommendation and Targeting}

We quantify the impact of the current RecSys on overall engagement and then investigate how this varies across goods. First, we compare the engagement under different reasonable recommendation algorithms to the currently deployed one. This allows us to quantify the overall gains in platform-wide engagement as a result of the current RecSys. Second, we estimate the impact of the RecSys in driving engagement for specific goods and quantify the incremental value of targeting across the distribution of goods.

The outcomes we consider allow us to quantify the firm value of the personalization enabled by the current RecSys.  In general, engagement and diversity are well known to correlate to long-term user engagement and retention, and so characterizing the effect of personalized recommendations on engagement is important for key firm objectives \citep{anderson2020algorithmic, bibaut2024learning, wang2022surrogate}. Although increased retention should reflect greater consumer welfare, our reduced-form model focuses on the impact of personalized recommendations on choice.  Since the underlying mechanisms that drive this can have different welfare interpretations, we refrain from making precise claims regarding the impact of the RecSys on consumer welfare.

\subsubsection{Value of the Current RecSys}\label{sec:value_of_recsys}

Since we have separately isolated the effect of recommendations from the baseline preferences of users, we can credibly simulate the impact of different recommendation systems on choices. We consider the impact of these different recommendation policies on two key variables: (i) overall engagement -- how much incremental consumption is generated -- and (ii) the diversity of goods that get consumed. Although overall engagement is the most relevant measure for the platform as it provides a measure for how much the RecSys is helping users find relevant goods, diversity is also important. This is because consumption diversity is beneficial in the long-run: it enables users to benefit from a broader set of goods and has been shown to correlate with higher long-run satisfaction \citep{anderson2020algorithmic, chen2024impact}.

We consider two measures of good-level diversity in consumption, where $s_{x}$ denotes consumption share of good $x$:
\begin{align}\label{eq:diversity}
\mathrm{HHI} &= \sum_{i=1}^{N} s_i^2, \quad
G = \frac{1}{2N} \sum_{i=1}^{N} \sum_{j=1}^{N} |s_i - s_j|
\end{align}

The Herfindahl–Hirschman Index (HHI) captures absolute concentration, as it places greater weight on goods with large exposure. It effectively measures how many goods receive meaningful viewership: as the consumption share concentrates on fewer goods, HHI rises toward one.
The Gini coefficient, in contrast, captures relative inequality in exposure. A higher Gini indicates that exposure is more unevenly distributed, even if many goods are still being viewed. Conceptually, HHI summarizes how broad the set of consumed goods is, while Gini describes how balanced exposure is within that set.

We compute \textsc{Engagement} and consumption diversity under the current RecSys compared to when $\mathcal{C}_{it}$ is generated using the following alternative algorithms:
\begin{enumerate}
\item \textbf{Random Recommendations} - $\mathcal{C}_{it}^{\text{RANDOM}}$: Every user gets a set of recommendations chosen uniformly at random from $\mathcal{M}$.
\item \textbf{Popularity Recommendations} - $\mathcal{C}_{it}^{\text{POPULAR}}$: Every user gets the same set of recommendations that are the top N goods in terms of market share in $\mathcal{M}$.
\item \textbf{Matrix Factorization Recommendations} - $\mathcal{C}_{it}^{\text{FACTOR}}$: The recommendations are computed using a matrix factorization algorithm \citep{koren2009matrix}. This captures the ``state of the art" approach to computing recommendations from 2015 as discussed in \cite{gomez2015netflix}.\footnote{We employ a vanilla SVD matrix factorization approach, without tuning or hyperparameter optimization. This is intended to provide a reasonable baseline for personalized recommendations, but does not represent a specific RecSys previously employed by Netflix.}
\end{enumerate}

To perform these counterfactual estimates, we simulate each alternative RecSys by replacing the set of recommended goods according to the relevant algorithm, keeping the number of recommendations each user receives constant. Our implementation takes into account several complexities of the actual good interface and business logic. Specifically, we model the effects of recommendations via a separate parameter for each of four different locations on the page the recommendations are present. This accounts for the fact that recommendations in certain locations (e.g., as described in Section \ref{sec:institutional_details}) are more salient and influential. When constructing counterfactual recommendations, we also disallow recommending goods that a user has already watched, as an approximation of the platform's discovery-oriented recommendations. Although changes to the mapping from predicted utilities to page positioning can significantly affect the resulting engagement and diversity metrics for a counterfactual recommendation algorithm, they do not substantially affect the relative ordering, and our approach closely follows the procedure employed by the RecSys.

The results of the comparison across algorithms is summarized in Table \ref{tab:delta_engagement_diversity}, reported in terms of percent deviations from the current RecSys. 

\begin{table}[ht!]
\centering
\caption{Engagement and Diversity under Counterfactual RecSys Algorithms}
\label{tab:delta_engagement_diversity}
\begin{tabular}{l c | cc}
\toprule
\textbf{Model} & \textbf{$\Delta$ \textsc{Engagement} (\%)} & \textbf{$\Delta$ Gini (\%)} & \textbf{$\Delta$ HHI (\%)} \\
\midrule
Current RecSys & - & - & - \\
Random & $-16$ & $-0.7$ & $-2.5$ \\
Popularity & $-12$ & $-0.1$ & $+42.5$ \\
Matrix Factorization & $-4$ & $+1.1$ & $+37.5$ \\
\bottomrule
\end{tabular}
\caption*{\footnotesize\textsc{Notes}: This table reports percent deviations from the currently deployed recommendation system (``Current RecSys") for overall engagement and two consumption-diversity metrics. 
\textsc{Engagement} is the probability of any play in a period as defined in Equation~\eqref{eq:engagement}. The Gini and HHI values are computed using Equation \eqref{eq:diversity}.}

\end{table}

First, these results imply that the current algorithm leads to large improvements in engagement relative to each of the benchmark algorithms.  Reverting to random, popularity, or matrix factorization algorithms would reduce engagement by 16\%, 12\%, and 4\%, respectively. These gains are quantitatively large and economically meaningful---e.g., internal estimates would imply that reverting back to matrix factorization recommendations would lead to similar monetary losses as the estimates from \cite{gomez2015netflix} indicate reverting back to popularity from matrix factorization would induce in 2015.\footnote{These estimates are likely lower-bounds, given that our implementation of the recommendation bonus implicitly incorporates the trust that the users have with the RecSys, which would likely be lower under these counterfactual policies.} This improvement highlights that, despite recent concerns in the RecSys community that modern algorithmic improvements are providing little gain relative to traditional approaches such as matrix factorization \citep{ferrari2019we}, the latest algorithmic improvements applied at one of the most prominent deployments of RecSys are leading to meaningful gains.

Second, we report both the HHI and Gini Index deviations from moving from the current RecSys to the alternative recommendation policies. We expect that random recommendation should lead to higher diversity, since by design it provides exposure to a wider set of goods. In line with this, we find that random recommendations lead to higher good diversity across both measures than the other recommendation policies. However, the difference is not dramatic---moving from the current RecSys to random recommendations leads to only a 0.7\% and 2.5\% reduction in these metrics, but with a large increase in engagement. In contrast, both matrix factorization and popularity-based recommendation policies lead to dramatic increases in concentration -- 37.5\% and 42.5\%, respectively -- as well as a +1.1\% and -0.1\% change in the Gini coefficient. These patterns indicate that both algorithms would substantially narrow the set of goods receiving meaningful viewership. The popularity-based policy concentrates exposure on a small group of equally dominant goods, whereas matrix factorization further amplifies inequality within this group.

Overall, these results show that the current RecSys meaningfully improves engagement while maintaining relatively high consumption diversity. In contrast, traditional matrix-factorization and popularity-based algorithms would lead to large reductions in engagement and concentrate viewership on a much narrower set of goods. This underscores that, despite the concerns in \cite{ferrari2019we}, modern improvements to recommendation algorithms are both helping match users with their preferred goods and resulting in improved consumption diversity which is important for sustaining a broader good ecosystem.

\subsubsection{Good-Level Value of Recommendation}

We now turn to understanding the value of recommendation for \textit{particular} goods by characterizing the incremental consumption for a good that is generated by the recommendation as well as the value of targeting in the effectiveness of recommendations. To begin, we define the potential outcomes for a particular user $i$ and good $j$ as follows:
\begin{align*}
Y_{ij}(1)&: \text{ consumption probability if good } j \text{ is recommended to user } i \\
Y_{ij}(0)&: \text{ consumption probability if good } j \text{ is \emph{not} recommended to user } i
\end{align*}

\begin{figure}[H]
\centering
    \caption{Good-level Recommendation Outcomes}\label{fig:good_treatment}
    \begin{subfigure}[b]{0.49\textwidth} 
        \caption{ATE}\label{fig:hist_ATE}
        \includegraphics[width=\textwidth]{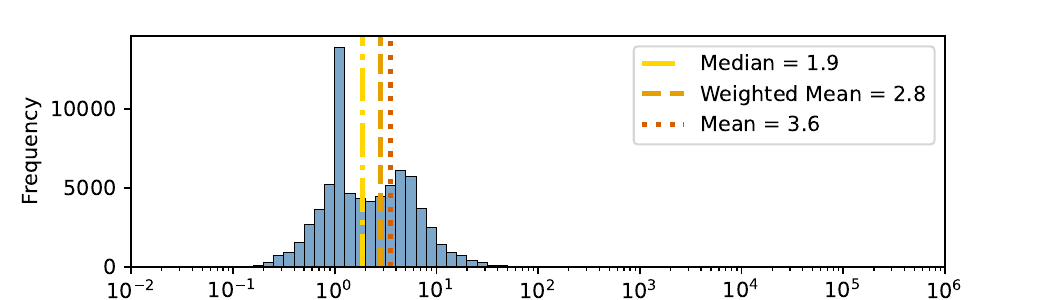}
    \end{subfigure}
    \begin{subfigure}[b]{0.49\textwidth}
        \centering
        \caption{ATT}\label{fig:hist_ATT}
        \includegraphics[width=\textwidth]{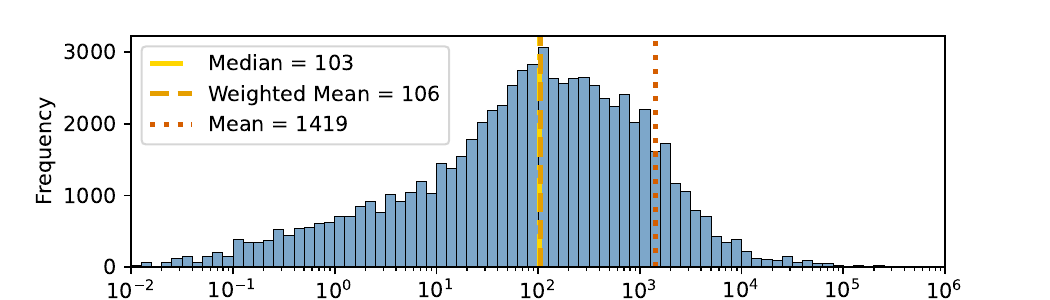}
    \end{subfigure}
    \begin{subfigure}[b]{0.49\textwidth}
        \centering
        \caption{$\bar{Y_{j}}^{T}(1) - \bar{Y_{j}}(0)$}\label{fig:hist_LHS}
        \includegraphics[width=\textwidth]{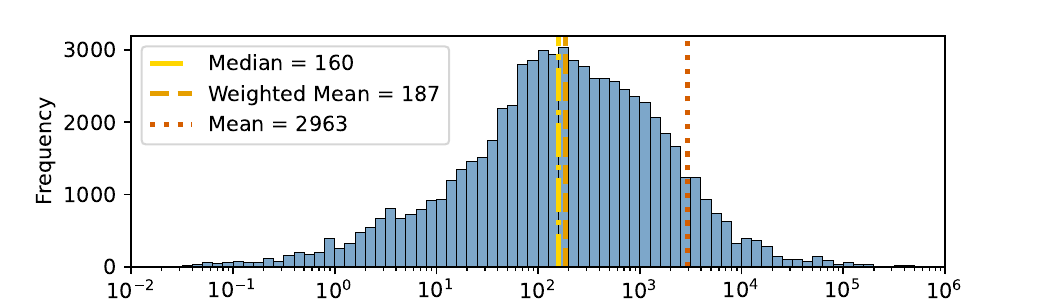}
    \end{subfigure}
    \begin{subfigure}[b]{0.49\textwidth}
        \centering
        \caption{$B_j$}\label{fig:hist_Bj}
        \includegraphics[width=\textwidth]{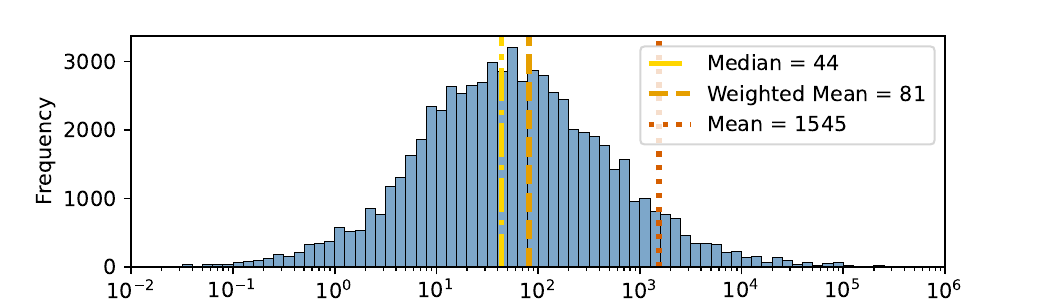}
    \end{subfigure}
    \begin{subfigure}[b]{0.49\textwidth}
        \centering
        \caption{$R_j$}\label{fig:hist_Rj}
        \includegraphics[width=\textwidth]{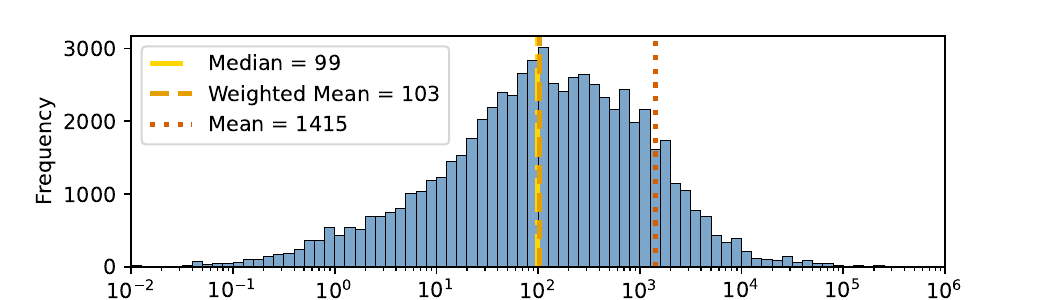}
    \end{subfigure}
    \caption*{\footnotesize\textsc{Notes}: We plot the distributions of each of the terms in \eqref{eq:decomposition}. To ease interpretation, we divide each through by $\bar{Y}_{j}(0)$. The different panels represent the following: (a) ATE$_j$ (average treatment effect of recommending good $j$ to a random user), (b) ATT$_j$ (average treatment effect on the treated, i.e., users who were actually targeted with $j$), (c) $\bar{Y}_{j}^{T}(1) - \bar{Y}_{j}(0)$ (difference between targeted-with-recommendation and population-without-recommendation probabilities), (d) $B_j$ (selection term), and (e) $R_j$ (targeting term), as defined in \eqref{eq:decomposition}. For each term, we report the median, baseline observation-weighted average, and mean across all goods.}

\end{figure}
Thus, using our model we can simulate both potential outcomes. To estimate $Y_{ij}(1)$ for each user $i$, we force good $j$ to be in each user's recommended set\footnote{If $j  \in \mathcal{C}_{it}$ originally, then we leave the rest of $\mathcal{C}_{it}$ same. If $j \notin \mathcal{C}_{it}$, then we take the average consumption probability of $N$ random goods to be replaced with good $j$ in  $\mathcal{C}_{it}$.} and to estimate $Y_{ij}(0)$ we remove good $j$ from $\mathcal{C}_{it}$. Thus, the causal effect of recommending good \(j\) to user $i$ is
\begin{align*}
    \tau_{ij} \equiv Y_{ij}(1) - Y_{ij}(0)
\end{align*}

\noindent We can then estimate the average treatment effect (ATE) of recommending good $j$ by
\begin{align*}
\text{ATE}_j = \frac{1}{I}\sum_{i=1}^{I}\tau_{ij}
\end{align*}
which captures the incremental consumption that would result if $j$ were recommended to a random user and the distribution of $\text{ATE}_j$ is presented in Figure \ref{fig:hist_ATE}. However, the ATE does not fully measure the value of the recommendation for specific goods as they are targeted by design towards users for whom the good would be a good fit. Thus, it is possible that some goods that have a low ATE of recommendations have a large effect for the users that they are actually targeted to. To characterize this, we consider the average treatment effect on the treated (ATT), using the realized recommendations as an (endogenous) treatment status. We characterize this as 
\begin{align*}
\text{ATT}_j = \frac{\sum_{i:\, j \in \mathcal{C}_{it}}\tau_{ij}}{\sum_{i}\mathbb{1}\{j \in \mathcal{C}_{it}\}}
\end{align*} and we plot the distribution of this across goods in Figure \ref{fig:hist_ATT}.

The substantially larger value of ATT compared to ATE highlights the importance of targeting in driving the effectiveness of recommendations. Intuitively, recommended goods are more likely to be consumed since (i) they have higher baseline consumption probability for targeted users (\textit{selection}), (ii) there is a base increase from having any good show up in recommendations (\textit{exposure}), and (iii) targeted users have a larger differential responsiveness than an average user (\textit{targeting}). We decompose the relative importance of each of these components in driving consumption for recommended goods.

In order to characterize this, we define several additional terms. We define $\bar{Y}_{j}(d)$ as the average consumption probability in the population with ($d=1$) and without $(d = 0)$ recommendation and $\bar{Y}^{T}_{j}(d)$ as the average consumption probability in the set of targeted users (i.e., $j \in \mathcal{C}_{it})$ with ($d=1$) and without ($d=0$) recommendation. This allows us to express the difference between the consumption probability of a recommended good in the targeted set of users and the consumption probability of a counterfactually not recommended good for an average user as follows:
\begin{align}\label{eq:decomposition}
\bar{Y}_{j}^{T}\!(1)-\bar{Y}_{j}\!(0)
&=\underbrace{\bar{Y}^{T}_{j}\!(0)-\bar{Y}_{j}\!(0)}_{B_j~(\text{selection})}
+\underbrace{\bar{Y}_{j}\!(1)-\bar{Y}_{j}\!(0)}_{E_j\equiv\text{ATE}_j~(\text{exposure})}
+\underbrace{\big[(\bar{Y}_{j}^{T}\!(1)-\bar{Y}_{j}^{T}\!(0))-(\bar{Y}_{j}\!(1)-\bar{Y}_{j}\!(0))\big]}_{R_j\equiv\text{ATT}_j-\text{ATE}_j~(\text{targeting})}
\end{align}
We compute these terms for each $j \in \mathcal{M}$ and report the distribution as well as the median, mean, and observation-weighted\footnote{The observation-weighted average weights each good by the total number of times it is observed across user recommendation sets, which corresponds to a typical recommended good, and mitigates bias from infrequently-recommended long-tail content.} average across goods for each term in Figure \ref{fig:good_treatment}. We first note that exposure alone has a large effect, leading to a nearly three times increase in consumption probability relative to the baseline consumption probability $\bar{Y}_{j}(0)$. However, we decompose the relative importance of each term in driving the consumption difference, finding that it is driven by 51.3\%, 6.8\%, and 41.9\% from selection, exposure, and targeting, respectively. This highlights that while recommending a good does meaningfully impact its consumption probability, the impact of targeting is dramatically larger (nearly 7 times larger) than the exposure effect alone. Combined, 93\% of the consumption differences comes from the RecSys identifying users with higher baseline consumption probability and with higher incremental value from targeting.

\begin{figure}[H]
\centering
    \caption{Consumption Decomposition Across Different RecSys Algorithms}\label{fig:decomposition}
    \includegraphics[width=0.9\textwidth]{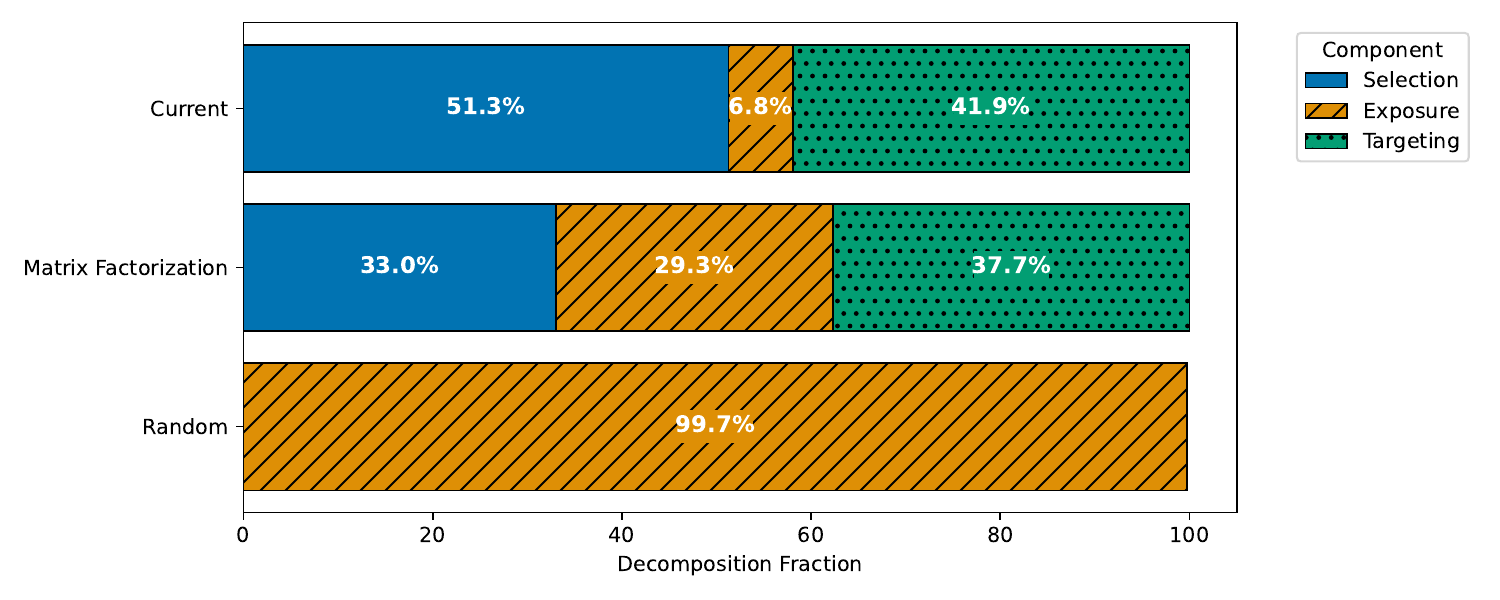}
    \caption*{\footnotesize \textsc{Notes}: This figure presents the relative importance of each of the terms in \eqref{eq:decomposition} -- selection ($B_{j}$), exposure ($E_{j}$), and targeting ($R_{j}$). The top, middle, and bottom rows present the relative importance for the current RecSys, the matrix factorization algorithm described in Section \ref{sec:value_of_recsys}, and the random recommendation algorithm described in Section \ref{sec:value_of_recsys}, respectively.}
\end{figure}

To contextualize these numbers, we provide two benchmarks that are displayed in Figure \ref{fig:decomposition}. The top row shows the decomposition for the current RecSys and the bottom two rows show the same decomposition for the matrix factorization and random algorithms discussed in Section \ref{sec:value_of_recsys}. The figure shows that random recommendations are dominated by the exposure effect which is due to the fact that, by construction, the random recommendation algorithm has no targeting. However, the decomposition for matrix factorization suggests that there is relative parity between each of the different components with exposure playing nearly as large of a role as both selection and targeting. This suggests that the results of increased engagement under the current RecSys from Section \ref{sec:value_of_recsys} are likely due to effective targeting and that the relative magnitude of targeting would not necessarily be as large even under other popular RecSys algorithms such as matrix factorization.

\begin{figure}[H]
\centering
    \caption{Targeting Responsiveness Heterogeneity}\label{fig:ratio_heterogeneity}
    \begin{subfigure}[b]{0.49\textwidth} 
        \caption{Density across Popularity vs $R_j$}\label{fig:ratio_popularity}
        \includegraphics[width=\textwidth]{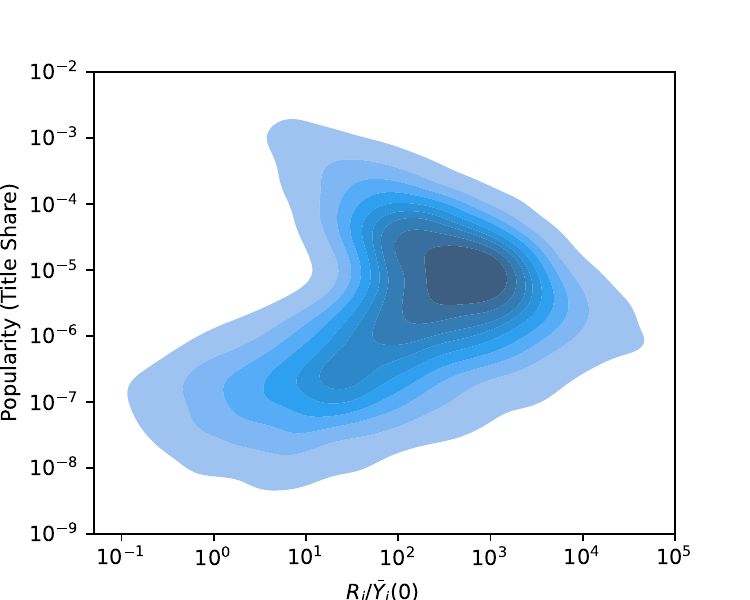}
    \end{subfigure}
    \begin{subfigure}[b]{0.49\textwidth}
        \centering
        \caption{Density across Good Category vs $R_j$}\label{fig:ratio_category}
        \includegraphics[width=\textwidth]{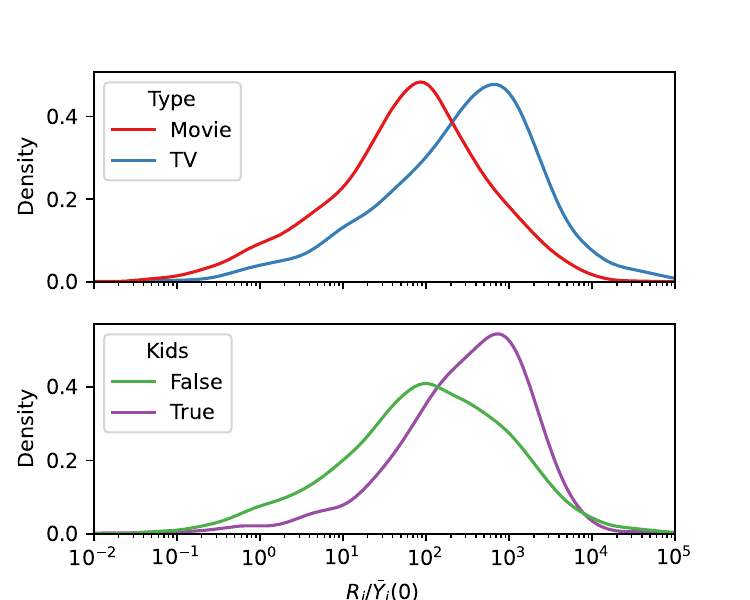}
    \end{subfigure}
    \caption*{\footnotesize\textsc{Notes}: We plot the kernel densities of $R_j$ (targeting) by (a) baseline popularity ($\bar{Y}_{j}(0)$) and (b) good categories (TV vs.\ Movie; Kids vs.\ Non-Kids). Higher $R_j$ indicates larger incremental responsiveness among targeted users beyond the average exposure effect. Category assignments are derived from internal metadata.}

\end{figure}

Finally, in Figure \ref{fig:ratio_heterogeneity} we explore heterogeneity in targeting ($R_{j}$) under the current RecSys---the most direct measure of targeting effectiveness for incremental consumption from recommendations---across observables based on baseline popularity ($\bar{Y}_{j}(0)$) and good category. This comparison allows us to ascertain whether niche goods gain the most from recommendations as well as which categories it is most important for. Figure \ref{fig:ratio_popularity} shows that targeting is most impactful for moderately popular goods. The most popular goods benefit less from targeting because of their extremely broad appeal, while the smallest goods may be too niche to target effectively, and mid-size goods represent a sweet spot of appealing to a sizeable but specific subset of users who may not discover the goods without recommendation. Additionally, we explore variation along other variables where we may expect heterogeneity: whether goods are TV or movies and whether a good is labeled as Kids or non-Kids. Figure \ref{fig:ratio_category} shows that TV benefits slightly more from targeting than movies, and Kids goods benefits more from targeting than non-Kids goods.

\subsection{The Incremental Demand for Goods}\label{subsec:incremental_demand}

The next application of the model that we consider is to estimate the incremental demand for particular types of goods, including new goods. Estimating this quantity is important for various platform applications, such as catalog optimization or scheduling of releases. 

We showcase how one can use the model to estimate these quantities, but due to confidentiality of these numbers do not report the actual estimates. For existing goods, we denote $\mathcal{M}_c$ as the set of existing goods we want to evaluate and define the incrementality of this set of goods as follows:
\begin{align*}
\textsc{Incrementality}^{EXISTING}(\mathcal{M}_c) = \sum\limits_{i}\widehat{\textsc{engagement}_{it}}(\mathcal{M}) - \widehat{\textsc{engagement}_{it}}
((\mathcal{M} \setminus \mathcal{M}_{c}))
\end{align*}

Intuitively, this allows us to express how much additional viewership this set of goods brings, accounting for substitution to other goods. In addition to measuring the incremental demand of existing goods, we use a similar definition for the incremental value of a new good ($\mathcal{M}_{n}$): 
\begin{align*}
\textsc{Incrementality}^{NEW}(\mathcal{M}_n) = \sum\limits_{i}\widehat{\textsc{engagement}_{it}}(\mathcal{M}_{n} \cup \mathcal{M
}) - \widehat{\textsc{engagement}_{it}}
(\mathcal{M})
\end{align*}
There are two key challenges with estimating these quantities. First, we need to simulate recommendations for goods that aren't currently available on the platform. Second, we cannot use our primary model for new goods as it relies on endogenously learned good characteristics.

For the first challenge, we can rely on the utility-based approach for imputing counterfactual recommendations -- described in Appendix \ref{sec:impute_recs} -- that uses draws from the predicted user-specific utilities.\footnote{We rely on the utility-based estimate procedure since changing the catalog would mean that the recommendation model is based on recommendations under a different catalog which would likely produce incorrect recommendations, especially for new goods.}  For the second challenge, we provide an extension of the model that relies on an ``exogenous" characteristic space that allows us to simulate counterfactual engagement for new goods.\footnote{This is a primary advantage of characteristic-space approaches to demand \citep{petrin2002quantifying}.}

\noindent \textbf{Model Extension: Exogenous Good Embeddings.} We therefore consider an extension of our model that replaces the endogenously learned good-level embeddings with an exogenous set of good characteristics. This allows us to credibly estimate counterfactual engagement for new goods. The empirical challenge is in constructing these good characteristics as this historically has prevented reasonable models of movie demand. We rely on good-level embeddings developed internally at Netflix that provide a high-dimensional representation of a good and capture many hard to directly measure characteristics.\footnote{Recent examples in economics that rely on embeddings in lieu of traditional characteristics spaces in demand models include \cite{quan2019extracting, toubia2019extracting, chen2020studying, bach2024adventures, magnolfi2025triplet, compiani2025demand}.} Importantly, these embeddings are trained based on pre-consumption characteristics and are primarily driven by human tags associated with each of the goods that are combined with observable characteristics, such as cast and genre, to produce a high-dimensional representation of existing and new goods that are not available on the platform.

\begin{figure}[H]
\centering
    \caption{Value of Data for Model Performance}\label{fig:vod_embeddings}
    \includegraphics[width=0.7\textwidth]{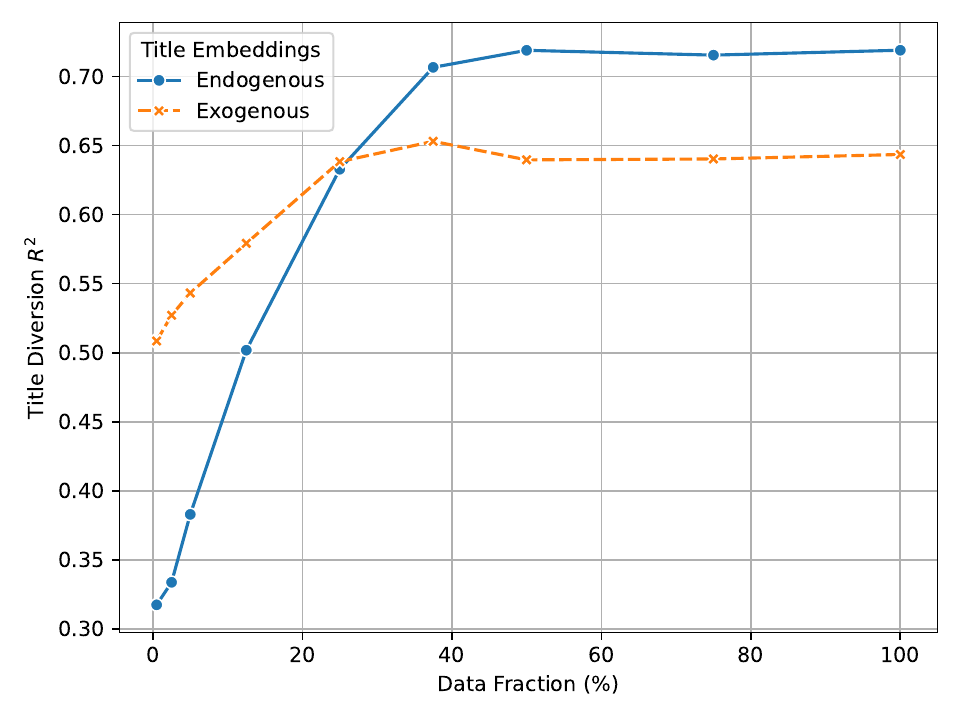}
    \caption*{\footnotesize\textsc{Notes}: This figure presents the value of data for the performance of the endogenous and exogenous embeddings models. The x-axis indicates the fraction of the 2 million users that we estimate the model on and the y-axis indicates the $R^2$ from the validation measures in Section \ref{sec:model_validation} for the resulting model estimates. Each data point is computed as the average over 4 runs for a given model and user fraction. We evaluate performance over the following grid of user fractions: \{ 0.5\%, 2.5\%, 5\%, 12.5\%, 25\%,  37.5\%,  50\%,  75\%, 100\% \}.}

\end{figure}

We provide estimates of this model and its validation in Appendix \ref{sec:app_exog_embeddings}. We report the same set of validation exercises as we conduct for our baseline model in Section \ref{subsec:experimental_comparison} and find that the model performs reasonably well. In particular, we show that the characteristic space is rich enough to predict the empirical good viewership shares with $R^2=0.96$ and that it similarly has a strong $R^2 = 0.65$ when comparing the empirical and model-based diversion estimates. This shows that we can use this model to credibly measure the incremental value of good for a wide class of goods.

Since the quality of the endogenous embeddings model depends on the scale of data needed to learn the good representations, a natural question is how much of the difference between these two models is driven by the amount of data used. In order to better understand this, we re-estimate both  models using a fraction of users compared to our primary sample and characterize how their performance on the experimental benchmarks from Section \ref{sec:model_validation} varies with the amount of data we use to estimate them. Figure \ref{fig:vod_embeddings} reports the results, showing large but diminishing returns to additional data for the endogenous embeddings model. At the extreme, when we use data from only 1\% of users the validation $R^2$ is nearly 60\% worse than the baseline. The performance exceeds that of the exogenous embeddings model only when we include around 30\% of users as the exogenous embedding model is more robust to limited data but less able to take advantage of additional observations. Overall, this seems to suggest that the superior performance of the endogenous embeddings model relies on the scale of data that we employ.

\section{Conclusion}

In this paper we studied the problem of separately measuring the value of goods and recommendations on a large streaming platform, Netflix. A key aspect of our approach is to exploit exogenous variation in the RecSys to both separately identify the effects of recommendations and ``intrinsic" preferences on choices in a discrete-choice model as well as to provide model-free diversion ratios that we use to benchmark the performance of our model. We showcase how to apply our model both to quantify the effect of recommendation on engagement as well as to ascertain the incremental value of goods.

Our paper highlights the central role of recommendation in driving user choices in these markets and the large value of the current RecSys on Netflix in several ways. First, exploration experiments in the RecSys can be a powerful lever to learn substitution patterns across goods. Second, we show that compared to both non-personalized and previous state-of-the-art methods for recommendations the current RecSys drives a meaningfully large amount of additional engagement. Third, we decompose the effects of recommendations on individual goods into selection into who gets targeted, the exposure effect of a recommendation, and differential responsiveness of targeted users from the recommendations. We show that while exposure meaningfully increases consumption, effective targeting plays a quantitatively larger role and that this effect is most pronounced for mid-sized goods, as opposed to very niche or broadly appealing goods. Overall, our paper provides rich insights into how to measure the value of recommendations for platforms and their impact on consumer behavior.

\bibliographystyle{chicago}
\bibliography{refs}

\appendix
\counterwithin{figure}{section} 
\counterwithin{table}{section}

\section*{Appendix}

\begin{appendices}

\section{Imputing Counterfactual Recommendations}\label{sec:impute_recs}

\begin{figure}[H]
\centering
    \centering
    \caption{Counterfactual vs. Observed Change in Recommendation Rate for Boosted Goods}
    \begin{subfigure}[b]{0.5\textwidth}   
        \caption{Utility-based estimate}
        \includegraphics[width=\textwidth]{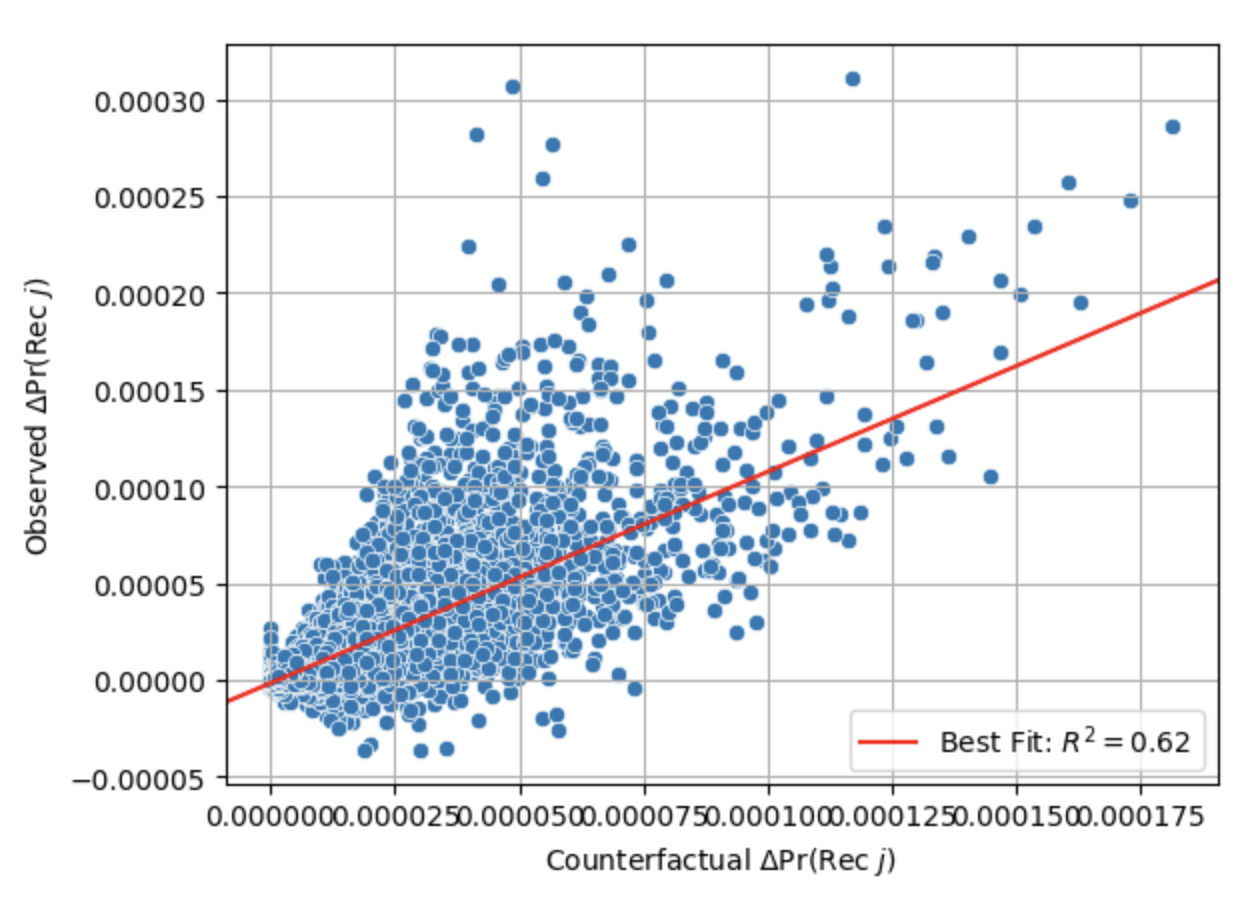}
        \label{fig:cf_recs_utility}
    \end{subfigure}%
    \begin{subfigure}[b]{0.5\textwidth}
        \caption{Recommendation model}
        \includegraphics[width=\textwidth]{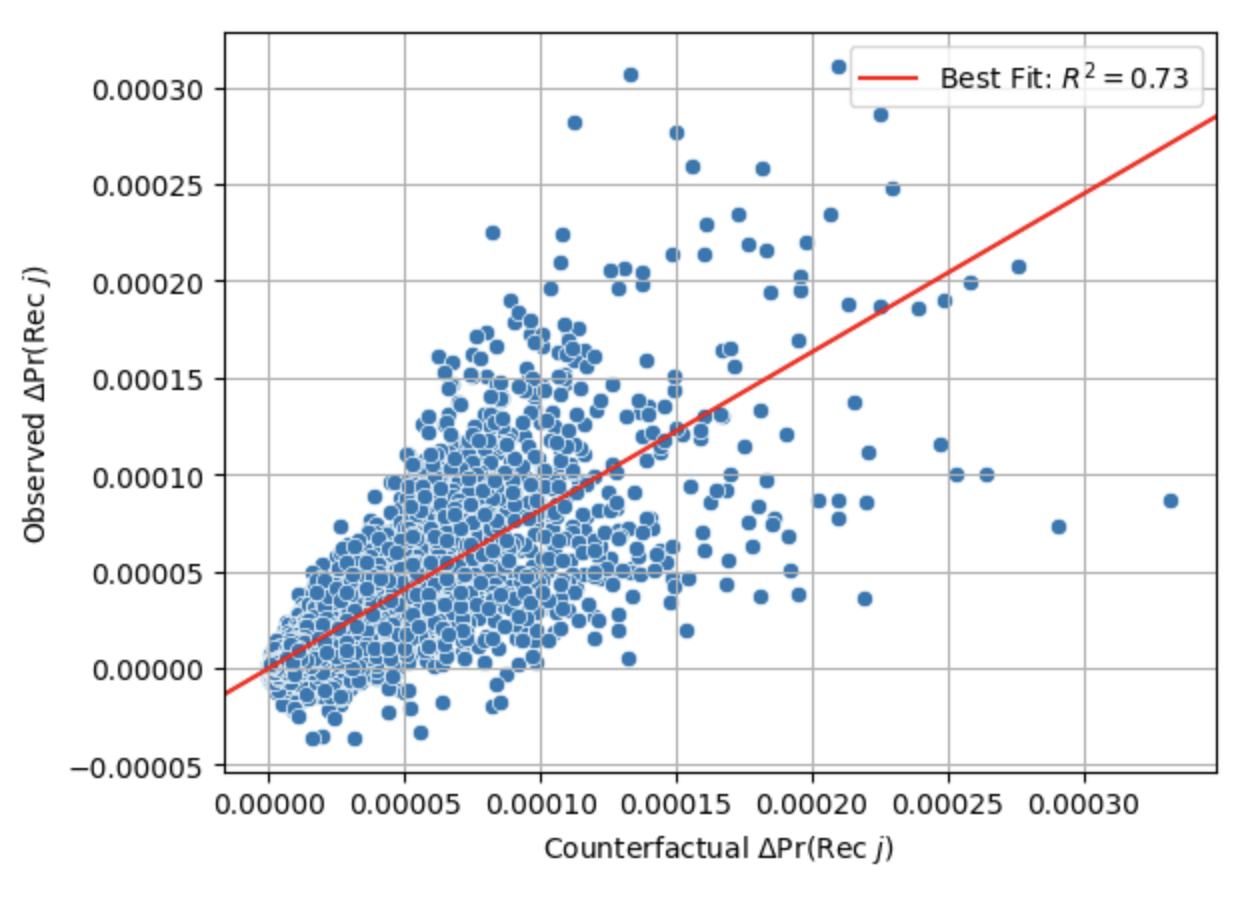}
        \label{fig:cf_recs_model}
    \end{subfigure}
    \label{fig:cf_recs}
    \caption*{\footnotesize\textsc{Notes}: For boosted goods in the salience experiments, we plot the change in recommendation rate predicted by (Panel a) the utility-based replacement rule and  (Panel b) the recommendation-model replacement against the realized change. Each point aggregates over users within an experimental arm and represents a single good. These procedures are used only to emulate recommendation replacement; demand estimation remains as specified in Section~\ref{sec:ccm}.}

\end{figure}

In this section we describe two methods for determining how to replace goods that appear in the recommendations. Since it is not possible for us to directly query the RecSys, we provide two procedures to do this imputation: a \textit{recommendation model} that estimates a separate model to learn the substitution patterns of the RecSys and a \textit{utility-based} estimate that relies only on the outputs from the demand model itself. Each of these procedures is useful for different applications -- for instance, the \textit{utility-based} procedure is important when we are considering counterfactual catalogs and the \textit{recommendation} model is important for ensuring we have an accurate as possible imputation of the recommendations during the intervention that we need for model validation.

The recommendation model approach involves training a model to predict which recommendations a user with a given history will receive. The structure of this model uses the same approach as our demand model for learning good embeddings $B_j$ and user embeddings $A_{it}$ as a function of watch history. However, rather than estimating a probability across a choice set, we estimate the probability of each good being recommended as:
\begin{align*}
\Pr(\textrm{Recommended } j \textrm{ to user } i \textrm{ at time } t) = \sigma(A_{it}B_j^\top)
\end{align*}

The utility-based procedure relies on the intuition that goods that appear in recommendations are those that have high predicted utility for user $i$. Thus, if a good is not allowed to appear in the recommendations, the next most substitutable good will also be one that has high predicted utility. While, of course, there is much more nuance in practice to the RecSys, our procedure follows this intuition by sampling goods using a softmax rule based on their predicted utility. Specifically, we consider the following replacement rule for $j \notin \mathcal{C}_{it}$:
\begin{align*}
\Pr(\text{Recommend } j \text{ to user }i) = \frac{\exp(u_{ij})}{\sum\limits_{k \in \mathcal{M}, k \notin \mathcal{C}_{it}}\exp(u_{ik})}
\end{align*}
We empirically validate both of these procedures by estimating the counterfactual recommendation rate in each treatment arm using our replacement rule. Then, we can subtract the baseline recommendation rate in the control group to obtain the increase in recommendations for boosted items. Finally, we can compare this counterfactual estimate to the observed change in recommendations for these items, as shown in Figure \ref{fig:cf_recs}. This validation demonstrates that we get reasonable estimates for the replacement rate for the boosted recommendations which are noisy but still sufficiently well-correlated for our purposes ($R^2 = 0.62$ under the utility-based estimate and $R^2 = 0.73$ under the recommendation model).

\section{Exogenous Embeddings Model Validation}\label{sec:app_exog_embeddings}

\begin{figure}[H]
\centering
    \caption{Exogenous Embedding Cross-Validation}\label{fig:exog_xv}
    \begin{subfigure}[b]{0.49\textwidth} 
        \caption{Good Share Cross-Validation}\label{fig:exog_shares_xv}
        \includegraphics[width=\textwidth]{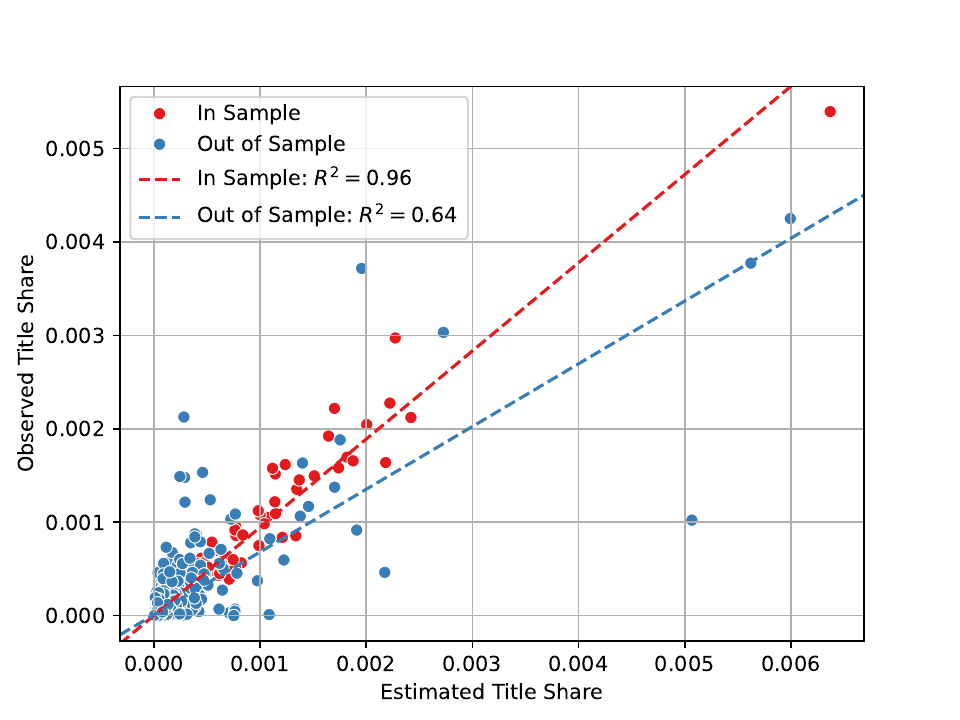}
    \end{subfigure}
    \begin{subfigure}[b]{0.49\textwidth}
        \centering
        \caption{Good Diversion Cross-Validation}\label{fig:exog_diversions_xv}
        \includegraphics[width=\textwidth]{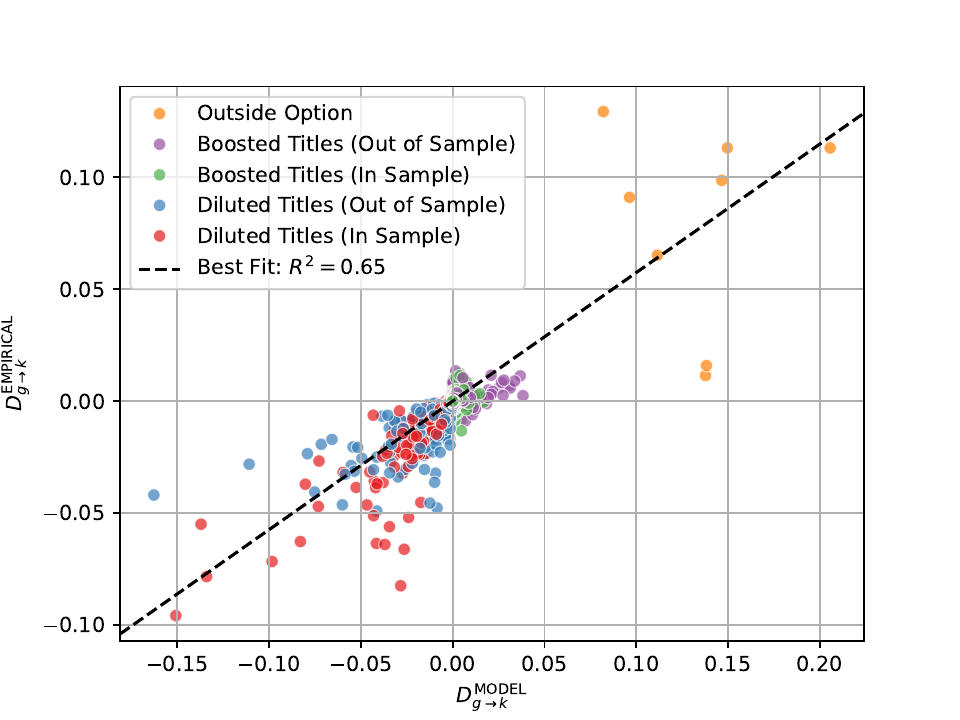}
    \end{subfigure}
    \caption*{\footnotesize\textsc{Notes}: This figure presents the validation of the exogenous embeddings demand model. Panel A presents the comparison between the model-based and empirical good shares. Panel B presents the comparison between the model-based and empirical good diversion. ``Out of sample" vs. ``in sample" is defined as follows. We randomly split the goods into two groups during model training. For the first group, we use the exogenous embeddings, and for the second group we learn endogenous embeddings according to the model in Section \ref{sec:ccm}. Then for evaluation, we replace all endogenous embeddings with exogenous embeddings, and label these goods as ``out of sample".}

\end{figure}

In this section we describe and validate an alternative version of the model from Section \ref{sec:ccm} where instead of endogenously learning the good embeddings, we rely on an exogenous characteristic space to represent goods. As discussed in the main text, a key challenge with historical demand models for movies is that choices can typically not be rationalized by standard characteristics such as cast and genre. As such, we rely on high dimensional representations of goods that are developed internally at Netflix and are constructed using pre-consumption human tags for each good.

We incorporate this into the model by imposing that $B_{j}$ is fixed through the estimation procedure according to this representation of goods. One challenge is that the dimensionality of the embeddings is still too high (256) for us to feasibly estimate in our model and so we include a two-layer, fully-connected multi-layer perceptron within the model to reduce the dimension from 256 dimensional to $d$ dimensional.

We validate this model using the same exact procedure as described in Section \ref{sec:model_validation} for the baseline model. Additionally, a concern is that, if the dimensionality reduction is built into the model training and rich enough, the model will effectively convert the exogenous embeddings into the endogenous embeddings, making it unusable for extrapolation to unseen goods (which is the primary purpose for the exogenous embeddings model). Thus, in addition to the exercises done in Section \ref{sec:model_validation}, we randomly split the goods into two groups during model training. For the first group, we use the exogenous embeddings, and for the second group we learn endogenous embeddings according to the model in Section \ref{sec:ccm}. Then for evaluation, we replace all endogenous embeddings with exogenous embeddings, and label these goods as ``out of sample".

The validation results are presented in Figure \ref{fig:exog_xv}. First, Figure \ref{fig:exog_shares_xv} shows that the in-sample fit for the market shares is fairly strong with $R^2=0.96$ and with the model able to rationalize the most watched goods. Additionally, it shows that the ``out of sample" $R^2=0.64$, indicating that the model does reasonably well at extrapolating to unseen goods. Second, Figure \ref{fig:exog_diversions_xv} shows that the model also does well at recovering the model-free diversion ratios with an $R^2 = 0.65$. Overall, we conclude that the model with the exogenous embeddings is able to sufficiently capture consumer choices for our purposes.

\end{appendices}

\end{document}